\documentclass[aps,a4paper,pre,twocolumn,groupedaddress]{revtex4}
\usepackage{graphicx}
\usepackage{subfigure}
\usepackage{amsmath,array}
\usepackage{amssymb,amscd,amsfonts}

\newcommand{\be}{\begin{equation}}
\newcommand{\ee}{\end{equation}}
\newcommand{\bea}{\begin{eqnarray}}
\newcommand{\eea}{\end{eqnarray}}
\newcommand{\noi}{\noindent}
\newcommand{\xt}{(\vec{x},t)}

\renewcommand{\vec}[1]{{ {\bf #1  }}}
\newcommand{\uvec}[1]{{ \hat{\bf #1}}}
\renewcommand{\r}[1]{\mathrm{#1}}

\begin{document}

\title{Stress response and structural transitions in sheared gyroidal and
  lamellar amphiphilic mesophases: lattice-Boltzmann simulations} 

\author{N\'elido Gonz\'alez-Segredo}
\email[Email: ]{nelido@amolf.nl}
\affiliation{FOM Institute for Atomic and Molecular Physics
(AMOLF), P.~O. Box 41883, 1009 DB Amsterdam, The Netherlands}
\author{Jens Harting}
\email[Email: ]{j.harting@ica1.uni-stuttgart.de}
\affiliation{Institut f\"ur Computerphysik, Pfaffenwaldring 27, 
D-70569 Stuttgart, Germany}
\author{Giovanni Giupponi and Peter V. Coveney}
\email[Email: ]{p.v.coveney@ucl.ac.uk}
\affiliation{Centre for Computational Science, Department of Chemistry, 
University College London, 20 Gordon Street, London WC1H 0AJ, United
Kingdom} 

\date{\today}

\begin{abstract}
We report on the stress response of gyroidal and lamellar amphiphilic
mesophases to steady shear simulated using a bottom-up lattice-Boltzmann 
model for amphiphilic fluids and sliding periodic (Lees-Edwards) boundary 
conditions. We study the gyroid {\em per se} (above the sponge-gyroid 
transition, of high crystallinity) and the molten gyroid (within such a 
transition, of shorter-range order). We find that both mesophases exhibit 
shear-thinning, more pronounced and at lower strain rates for the molten 
gyroid. At late times after the onset of shear, the skeleton of the 
crystalline gyroid becomes a structure of interconnected irregular tubes 
and toroidal rings, mostly oriented along the velocity ramp imposed by 
the shear, in contradistinction with free-energy Langevin-diffusion studies 
which yield a much simpler structure of disentangled tubes. We also compare the 
shear stress and deformation of lamellar mesophases with and without 
amphiphile when subjected to the same shear flow applied normal to the lamellae. 
We find that the presence of amphiphile allows (a) the shear stress at 
late times to be higher than in the case without amphiphile, and (b) the 
formation of rich patterns on the sheared interface, characterised 
by alternating regions of high and low curvature.
\end{abstract}


\maketitle

\section{Introduction}

The study of the response to shear in amphiphilic mesophases has 
been the subject of attention for numerical modellers only in recent 
years. The interest in the subject is sustained not only by the wide 
range of  applications in materials' science and chemical engineering, 
but also by the need to gain a fundamental understanding of the universal 
laws governing the self-assembly processes and competing mechanisms present.

Hitherto, studies have focused mainly on the structural changes induced 
by steady and oscillatory shear, near and far from critical points, in 
polymer systems~\cite{ZVELIN1,ZVELIN2,ZVELIN3,MOROZOV,MOROZOV2,BOEK}. 
The morphologies studied include cubic- and wormlike-micellar, lamellar and 
hexagonally-packed-tubular mesophases; more complex structures are the 
so-called bicontinuous mesophases, of which those liquid-crystalline of 
cubic symmetry have thus far been considered in far less detail. 

The amphiphilic gyroid~\cite{P3publon,P3} is a bicontinuous cubic liquid 
crystal consisting of multi- or mono-layer sheets 
of self-assembled amphiphile dividing two regions, each containing phases
which are mutually immisicible, e.g. aqueous and hydrocarbon species. 
These sheets or {\em labyrinths} 
form a triply periodic minimal surface (TPMS) whose unit cell is of cubic 
symmetry, has zero mean curvature, no two points on it are connected 
by a straight segment, and has no reflexion symmetries. Their skeletons, 
i.e. the locus bounded by the TPMS, for each immiscible phase,
form double (inter-weaving), chirally symmetric 3-fold coordinated lattices.
There are lyotropic~\cite{P3publon,P3} and thermotropic transitions between 
the gyroid and the microemulsion mesophase, the latter being a bicontinuous 
mesophase of short-range order. The morphologies in the crossover regions 
of these transitions show shorter-range order than the gyroid's and 
longer-range order than the microemulsion's, reasons for which they are 
termed `molten gyroids.'

Bicontinuous cubic mesophases of monoglycerides and the lipid extract
from archaebacterium {\em Sulfolobus solfataricus} have been found at
physiological conditions in cell organelles and physiological
transient processes such as membrane budding, cell permeation and the
digestion of fats \cite{MARIANI}. They can also be synthesised
for important applications in membrane protein crystallisation,
controlled drug release and biosensors \cite{MARRINK,LUZZATI}. 

The purpose of this paper is to report on the response to shear of 
gyroid (G), molten-gyroid (MG) and lamellar (L$_\alpha$) amphiphilic 
mesophases simulated using a bottom-up kinetic-theoretic model for fluid
flow. The model is based on a lattice-Boltzmann (LB) 
method, which has proved to be a modelling tool alternative to and more 
efficient and robust than sofisticated methods based on continuum equations. 
This LB method adheres 
to a {\it bottom-up} complexity paradigm~\cite{COVNOBEL} in the sense that 
it is simple, fully particulate and no hypotheses of desirable macroscopic 
behaviour are imposed on the microdynamics---yet, we have shown in the past 
its ability to simulate correct segregation kinetics for immiscible 
fluids~\cite{P2} and non-equilibrium self-assembly into amphiphilic 
mesophases~\cite{P3publon,P3}. Knowing that such a simple model is capable
of simulating these kinetic processes from a purely bottom-up dynamics, 
in this paper we investigate how hydrodynamic interactions couple with 
self-assembly and 
modify the stability and morphology of the mesophases. The novelty of our
work also rests in the model's capability to reproduce morphological 
transitions without having to assume a macroscopic, free-energy model, 
used in other LB methods \cite{ZVELIN3,XU} to compute the diffusive 
currents substantiating self-assembly.

In addition, since our method models amphiphilic molecules as point
dipoles---the simplest possible particulate model for an amphiphile---the 
rheological features emergent from it are expected to be universal for a broad 
range of amphiphilic systems. Finally, most of the numerical studies 
measuring the stress response of complex
fluids to shear reported in the literature deal with phase-segregating 
fluids on one side~\cite{XU}, and the more complicated polymeric~\cite{GROOT} 
and glassy systems on the other~\cite{SOLLICH}. In this respect, the present 
article stands somewhere in between these two.

Our paper is structured as follows. In the next section we briefly introduce 
the model and describe the boundary conditions for the imposition of shear. 
In Section~\ref{SHEARTHINNING} we report on simulation data and conclude that 
shear-thinning occurs for both G and MG mesophases, leading to a transition to 
a mesophase consisting of tubular and ring-like structures as the strain 
increases. In section~\ref{LAMELLAR} we reveal how the presence of amphiphile in 
lamellar mesophases induces the formation of rich interfacial patterns surviving 
shear and allows higher values of stress than in lamellar mesophases 
without amphiphile. Finally, we provide our conclusions in Section~\ref{CONCL}.

\section{The model and the Lees-Edwards boundary conditions}
\label{LEBC-SECTION}

We utilised an existing bottom-up lattice-Boltzmann 
(LB) model for amphiphilic fluids \cite{P3publon,P3}, extended to 
simulate shear flow by means of Lees-Edwards boundary 
conditions~\cite{LE}. The model is in turn based on an extension made
to the Shan-Chen bottom-up LB model for immiscible fluids to model 
amphiphilic-fluid flow, and employs 25 microscopic velocities, of speeds
0, 1 and $\sqrt{2}$, in three dimensions (D3Q25 
lattice)~\cite{BOGH,LGA_D3Q25}. The model uses a BGK (relaxation-time)
approximation to the collision term of the Boltzmann equation for fluid 
transport, which allows us to simulate, for large enough lattices~\cite{SUCCI}, 
the Navier-Stokes (NS) momentum-balance equation in the bulk of each 
immiscible fluid species, namely ``oil'' (or ``red", `r') and ``water'' 
(or ``blue", `b'). The model allows the simulation of correct phase-segregation 
kinetics in the absence \cite{P2} and presence \cite{P3} of a third, 
amphiphilic (surfactant-like, ``s") dipolar species. The model 
controls the inter-particle forces between r, b, and s species via
coupling parameters ($g_\r{br}, g_\r{bs}, g_\r{ss}$), and transients
are controlled via relaxation times for densities ($\tau^\r{b}, \tau^\r{r}, 
\tau^\r{s}$) with an additional relaxation time for the orientation of the
amphiphile dipoles ($\tau^\r{d}$).  In addition, the model simulates the 
nonequilibrium self-assembly and relaxation dynamics of sponge (L$_3$) 
and gyroid mesophases~\cite{P3publon,P3}. The gyroids that it simulates
show rigidity, arising from their crystalline ordering, which decreases as 
the concentration of amphiphile is reduced; indeed,
a lyotropic transition causes the correlation length to decrease towards that 
of a sponge mesophase through a molten-gyroid state. This idea is 
central to the work we present here: we shall see that the mesophase's 
crystalline ordering enhances its stress response; indeed, we find 
shear-thinning to occur at higher strain rates for gyroids than for sponges. 

The Lees-Edwards boundary conditions (LEBC) were originally proposed by 
Lees and Edwards in the context of molecular dynamics simulations~\cite{LE}. 
They showed that these boundary conditions would give rise to a desired 
linear, wedged velocity profile whilst avoiding the troublesome spatial 
inhomogeneities appearing when solid walls are used to induce the shear 
flow~\cite{WAGNER}. A particular realisation of the LEBC on a cartesian 
simulation box $[0,N_\r{x}]\times[0,N_\r{y}]\times[0,N_\r{z}]$ is established 
by letting the periodic images, at $N_\r{x}<x\le 2N_\r{x}$ and
$-N_\r{x}\le x\le 0$, move parallel to unit vectors $\pm{\uvec z}$, 
respectively, both with speed $U$. The LEBC, in their original, molecular
dynamics form, are expressed as a Galilean transformation on the position 
$(x,y,z)$ and velocity $(\xi_\r{x},\xi_\r{y},\xi_\r{z})$ co-ordinates of a 
molecule, as follows
\bea
   x^{\prime} &\equiv& 
   x\,\r{mod}\,N_\r{x}              \nonumber\\
   y^{\prime} &\equiv&
   y\,\r{mod}\,N_\r{y}              \nonumber\\
   z^{\prime} &\equiv&
   \left\{
   \begin{array}{ll}
     (z + \Delta_\r{z})\,\r{mod}\,N_\r{z} & ,\,x>N_\r{x},\\
     z\,\r{mod}\,N_\r{z}                  & ,\,0\le x\le N_\r{x},\\
     (z - \Delta_\r{z})\,\r{mod}\,N_\r{z} & ,\,x<0,\\
   \end{array}
   \right .                        \label{LEBC-1}\\
   \xi_\r{x}^{\prime} &\equiv& 
   \xi_\r{x}                       \nonumber\\
   \xi_\r{y}^{\prime} &\equiv& 
   \xi_\r{y}                       \nonumber\\
   \xi_\r{z}^{\prime} &\equiv&
   \left\{
   \begin{array}{ll}
     \xi_\r{z} + U             & ,\,x>N_\r{x},\\
     \xi_\r{z}                 & ,\,0\le x\le N_\r{x},\\
     \xi_\r{z} - U             & ,\,x<0,\\
   \end{array}
   \right .                        \label{LEBC-2}
\eea
\noi where $\Delta_\r{z}\equiv U\Delta t$ is the image's shift at 
time $\Delta t$ after the onset of shear.

An implementation of the LEBC on our LB dynamics (LB-LEBC) differs 
from that used in molecular dynamics (MD-LEBC) in that the shift 
$\Delta_\r{z}$ is not in 
general a multiple of the lattice unit, as Wagner and Pagonabarraga have
pointed out~\cite{WAGNER}, and hence an interpolation scheme is
needed. This interpolation scheme streams the amphiphile dipoles 
${\vec d}({\vec{x}})$ and mass densities $n_k^\alpha({\vec{x}})$
located at position $\vec{x}$ on the shearing wall, where ${\vec c}_k$ 
is the relevant discrete molecular velocity, $k=1,\ldots,25$, for each 
(fluid and amphiphilic) species $\alpha$.

In our LB-LEBC, while the spatial displacement follows Eqs.~(\ref{LEBC-1}),
the velocity shift cannot be enforced by replacing the continuum 
velocity component ${\vec{\xi}}_\r{z}$ in Eqs.~(\ref{LEBC-2}) with the 
discrete microscopic speeds ${\vec c}_k\cdot\uvec{z}$, since the 
velocities ${\vec c}_k$ are constant vectors. Instead, this acceleration
is enforced on the macroscopic fluid velocity around which the 
local-Maxwellian distributes molecules at equilibrium and towards which 
the BGK scheme relaxes, similarly to how immiscibility forces are 
implemented~\cite{P2,P3}. This procedure guarantees that all accelerations 
in the fluid are ruled by the same BGK process, controllable via the shape 
of the distribution function and the relaxation-time parameter, including 
the acceleration due to the shearing walls. 

The MD-LEBC give rise, at steady state (late times), to a shear state which is 
Galilean-invariant, i.e. no particular plane in the system is favoured over
another. This is a {\it sine qua non} for any shearing method, and our method 
satisfies it too. As regards the unsteady, transient initial states, the 
MD-LEBC are unphysical since they cannot provide the molecular specificity 
(e.g. wall roughness) required in an atomistic approach 
to boundary effects, such as density layering and slip at wall. 
However, mesoscopic methods---LB is one of them---in general only describe low 
wavenumbers and frequencies, which means that, with respect to MD, 
(a) the atomistic detail of the shearing walls
    is largely coarse-grained and (b) the fluid structure and dynamics are
    much less sensitive to the atomistic detail of the walls. 
Since most boundary effects present in MD are absent in LB, the fact that the 
LE boundary conditions eliminate them does not pose a problem.
This should be taken with a caveat: our gyroidal mesophases melt when
    placed in a solid box, which means that the approach to equilibrium
    is sensitive to momentum transfer with the walls, and therefore the LE
    boundary conditions do not mimic shearing a mesophase in confinement.
    (To our knowledge, no bottom-up simulations have ever reported mesophase
    self-assembly in confinement.)  Rather, the LE boundary conditions in
    a LB model mimic shearing with walls which are far enough from the
    locations in the system where observables are probed such that microscopic
    boundary effects are absent.

Our LEBC implementation is embedded within an efficient parallel LB
algorithm~\cite{CPC} which allows us to employ large lattices and
hence reach the small Knudsen number limit where (a) regions away from
interfaces satisfy the incompressible NS equation in the limit of low
Mach numbers (Ma)~\cite{P2}, and (b) observables vary by less than 10\%
when the lateral lattice dimension is doubled. We have previously found 
that the lattice size guaranteeing condition (b) is $128^3$ for the 
parameters generating the mesophases investigated here~\cite{P3publon,P3}.

\section{Shearing gyroidal mesophases}
\label{SHEARTHINNING}

We sheared two gyroidal mesophases differing in the amount of 
amphiphile present and the value of the inter-amphiphile interaction
coupling parameter. Each of these structures was allowed to self-assemble
from homogeneous mixtures of oil, water and amphiphile using periodic 
boundary conditions. They have been appropriately characterised by probing 
direct and Fourier-space late-time snapshots of the density order parameter
$\phi\equiv\rho^\r{oil}-\rho^\r{water}$; more precisely, they correspond to 
gyroid (cf. Fig.~\ref{MORPH_G}(a)) and molten gyroid mesophases, as previously 
reported by us~\cite{P3publon,P3}. 

The common parameters used for both gyroids were oil and water densities 
flatly distributed in the range
$0<n^{(0)\mathrm{b}}=n^{(0)\mathrm{r}}<0.7$, coupling strengths
\mbox{$g_{\mathrm{br}}=0.08$}, \mbox{$g_{\mathrm{bs}}=-0.006$}, relaxation times 
$\tau^{\mathrm{b}}=\tau^{\mathrm{r}}=\tau^{\mathrm{s}}=\tau^{\mathrm{d}}=1$, 
and, for the amphiphile's dipoles, $\beta=10$ and $d_0=1$. 

Their differing parameters were surfactant densities, flatly 
distributed in the initial homogeneous mixture, in the ranges 
$0<n^{(0)\mathrm{s}}<0.9$ for the gyroid and $0<n^{(0)\mathrm{s}}<0.6$
for the molten gyroid, with coupling strengths $g_{\mathrm{ss}}=-0.0045$
for the gyroid and $g_{\mathrm{ss}}=-0.003$ for the molten gyroid. These 
values for the gyroid are 50\% higher than those for the molten gyroid. 

While the gyroid relaxes to a highly crystalline structure~\cite{TERAGYROID}, 
the molten gyroid shows both shorter-range order and stronger temporal 
fluctuations than the former~\cite{P3}. In order to obtain a sufficiently
relaxed molten gyroid as an initial condition for the shear, we took the 
structure as evolved up to time step 32\,500; regarding the gyroid, the 
time slice chosen was time step 15\,000. For practical reasons, instead of
letting the molten gyroid self-assemble starting from a homogeneous initial 
mixture, we upscaled a smaller molten gyroid, previously self-assembled 
using the same parameters on a $64^3$ lattice~\cite{P3}, to a $128^3$ 
lattice. Upscaling consisted in replicating identical copies of the system:
the periodic boundary conditions used to generate the $64^3$ system 
(a)~guarantee that the density field is smooth across the replica boundaries, 
yet, for this same reason, (b)~produce a
molten gyroid with an additional, undesirable long-wavelength fluctuation 
whose periodicity is half the lattice size. The amplitude of this undesired 
long-wavelength 
fluctuation relaxes in time to a vanishingly small value, fact which provides 
us with the $128^3$ mesophase we seek. We observed, however, that this 
relaxation takes place in less than 1\,000 time steps~\cite{TERAGYROID}, 
i.e. it is a fast transient which, therefore, does not affect the shear 
response at the late times that we are interested in. In other words, the 
late-time shear response is insensitive to a small perturbation in the initial 
condition. This allowed us to take the upscaled, unrelaxed structure as the
initial condition for the molten gyroid.

It is worth noting that we did not require an elongated aspect ratio 
for the lattices along the direction parallel to the translation of the 
shearing walls since spatial density fluctuations were much smaller 
than the lattice size. This is not the case when shearing phase-segregating 
fluids without an amphiphilic, growth-arresting species, as has been 
previously reported using LB lattices of up to \mbox{$128:128:512$} sizes 
and aspect ratio~\cite{JENS}.

\subsection{Stress response and transients}

Shear thinning is said to occur when the shear viscosity drops as the strain 
rate increases. For structured fluids such as those we study in this paper, the 
dynamic shear viscosity, $\eta$, is not expected to be a constant of the strain
rate $\dot\gamma\equiv\frac{1}{2}(\partial_{\r x}u_{\r z} + 
\partial_{\r z}u_{\r x})$ as is true of Newtonian fluids, for which, 
\be
P_{\mathrm{xz}}
=
\pm
2\eta\,\dot\gamma
\,,\qquad\eta\neq\eta(\dot\gamma)\,.
\label{NEWTON}
\ee
\noi Here $P_\r{xz}$ is one off-diagonal component of the pressure (or stress)
tensor, and the sign, by convention, indicates that the pressure is exerted by
the fluid element on the surroundings (`$+$') or from the latter on the former
(`$-$'), respectively. We adhere in this paper to the second case.
In our simulations, we apply the steady shear described in Section~\ref{LEBC-SECTION}, 
i.e., the shear is generated by the two image cells of the LB lattice located 
along the $x$-axis moving in opposite directions. As a consequence, 
$\partial_{\r x}u_{\r z}$ becomes the only non-vanishing component of the velocity 
gradient, which is also true for the  $P_\r{xz}$ component of the stress tensor (and 
$P_\r{zx}$, since the physical requirement that the vorticity,  
$W\equiv\frac{1}{2}(\partial_{\r x}u_{\r z} - \partial_{\r z}u_{\r x})$, remains 
upper bounded requires the stress tensor to be symmetric).

As we have likewise done previously while computing diagonal components 
of the pressure tensor~\cite{P2,P3publon,P3}, here we measured $P_{\mathrm{xz}}$ 
from its definition as the sum
of a kinetic term plus a virial mean-field term accounting for
interactions and giving rise to non-ideal gas behaviour, namely,
\bea
	{\mathsf P}({\mathbf x})
	&\equiv&
	\sum_{\alpha}\sum_k\rho_k^{\alpha}({\mathbf x}) 
	({\mathbf c}_k-\vec{u(x)}) 
	({\mathbf c}_k-\vec{u(x)})                       \nonumber\\
	&+&
	\frac{1}{4}\sum_{\alpha,\bar\alpha} g_{\alpha\bar\alpha}
	\sum_{{\mathbf x}'} \Big[ \psi^{\alpha}({\mathbf x})
	\psi^{\bar\alpha}({\mathbf x}') + \psi^{\bar\alpha}({\mathbf
	x}) \psi^{\alpha}({\mathbf x}') \Big]
	\times                                           \nonumber\\
	&& (\mathbf{x-x'})(\mathbf{x-x'})		 \,,
	\label{PRESSURETENSOR}
\eea
\noi where $\psi$ has the form $\psi\equiv1-\exp[-n(\vec{x})]$, 
which saturates at high density values in order to avoid unbounded
interparticle forces whilst reproducing a meaningful equation of 
state~\cite{P3}. Since the interaction matrix 
$\{g_{\alpha\bar\alpha}\}$ is symmetric with all diagonal elements 
identically zero, and only nearest neighbour interactions are being 
considered, the virial term reduces to
\be
	\frac{1}{2}\sum_{\alpha\neq\bar\alpha} g_{\alpha\bar\alpha}
	\sum_{k}\psi^{\alpha}({\mathbf x})
	\psi^{\bar\alpha}({\mathbf x}+{\vec c}_k) 
	{\vec c}_k{\vec c}_k		 \,. 
\ee
\noi 
In the incompressible, low Mach number limit, our LB model reproduces the 
NS equation away from interfaces~\cite{BOGH,SUCCI}, which describes a
Newtonian fluid with a viscosity being a well known function of the
relaxation time. The presence of an interface, characterised by an interfacial
tension and a bending rigidity, however, introduces anisotropies in the
fluid's stress tensor which can be accounted for by a tensorial effective 
viscosity. Since the interface may move, at a speed growing with the strain rate, 
these anisotropies can become unsteady. Our aim is then to measure how this
viscosity evolves with the strain and the strain rate.

In order to probe the function $\eta=\eta(\dot\gamma)$ for both gyroidal 
mesophases, we measured $P_{\mathrm{xz}}$ for a number of different
applied shear rates. The chosen values for $U$ were such that they 
remained within the incompressibility limit, i.e. small compared to the 
speed of sound on the D3Q25 lattice, $c_\mathrm{s}=3^{-1/2}\approx0.58$. 
Values chosen were $U=0.05,0.10,0.15,0.20$, corresponding to Mach 
numbers $\mathrm{Ma}\equiv U/c_\r{s}=0.086, 0.17, 0.26, 0.34$, respectively. 
All observables we report in this paper are spatial averages,
at least on $x=\r{const.}$ planes where a simple fluid under the same 
shear would show translational symmetry for the velocity field, i.e., 
perpendicular to the velocity gradient. Since, for reasons of computational
cost, we do not perform averages over the seed used to generate the
pseudo-random initial configuration mimicking a homogeneous ternary
mixture, we do not provide error bars around averages.

\begin{figure}[!htb]
\includegraphics[angle=-90,width=10.5cm]{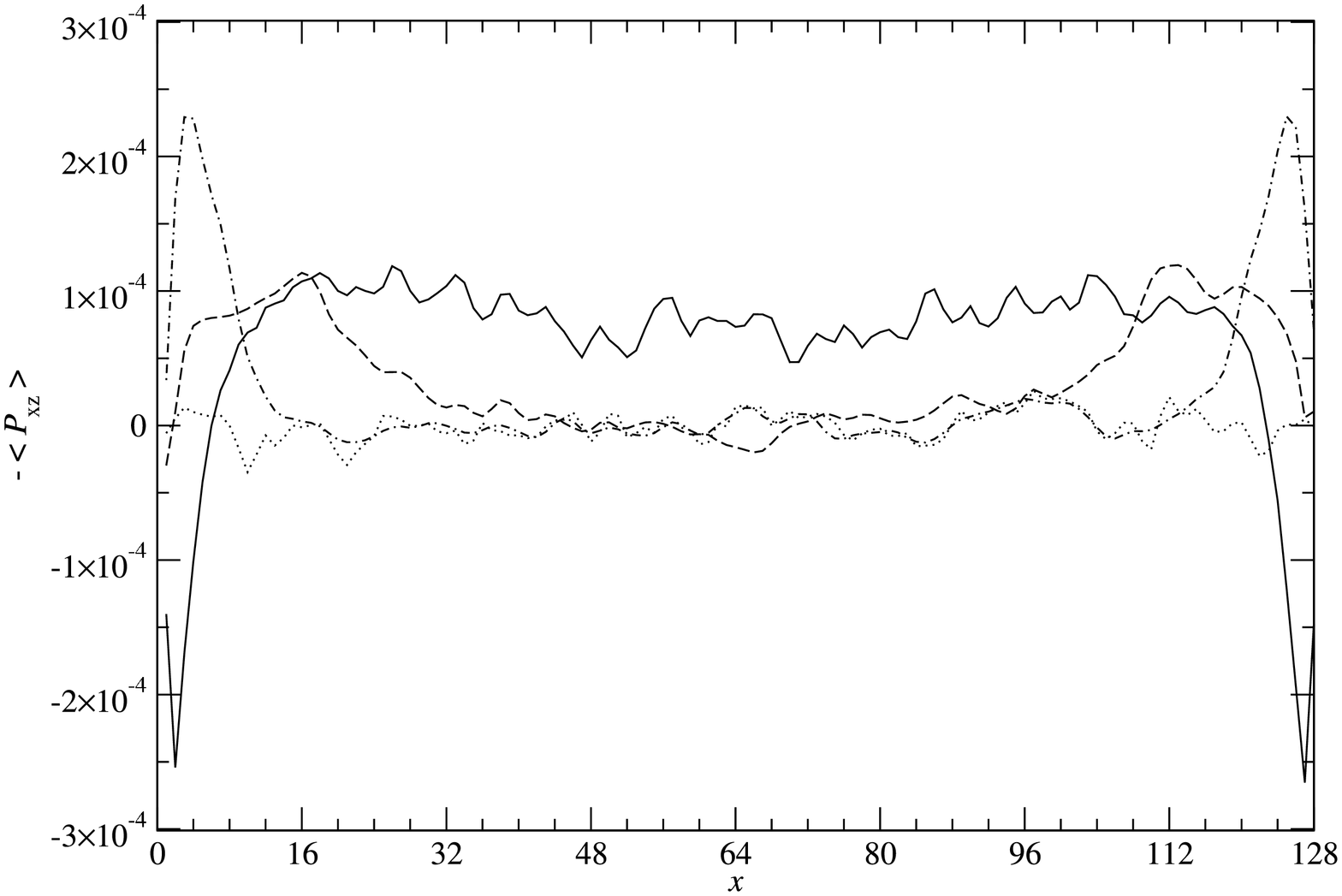}
\caption{\footnotesize\linespread{0.9}
  Shear stress response of a gyroid mesophase along the
  direction of the velocity gradient. As initial condition, we have
  taken a gyroid on a $N_{\r x}N_{\r y}N_{\r z}=128^3$ cubic lattice
  at time step $t=15\,000$ of self-assembly \cite{P3publon,P3}. The
  Lees-Edwards walls move with  
  speed $U=0.10$ ($\r{Ma}=0.17$). For each $x$ coordinate, the
  original field has been averaged on the plane 
  $[1,N_{\r y}]\times[16,N_{\r z}-16]$, where the excluded interval 
  on the $z$-axis accounts for
  wrapped-round densities. Standard errors of the averages are about
  $6\times10^{-8}$ throughout, and are not shown. Each line represents
  the response at $\Delta t$ time steps after the start of steady
  shear:  $\Delta t=0$ (dotted line), $\Delta t=100$ (dash-dotted),
  $\Delta t=800$ (dashed) and $\Delta t=9\,000$ (solid), where the
  last is ca. the time at which the core (i.e., the plane $x=64$)
  fully responds. From the figure we can see that momentum transfer 
  decreases as it reaches the core from the walls. 
  Also, note that the stress inverts its sign at late times adjacent 
  to the boundaries, $|x-x_0|\le2$ ($x_0=0,\,128$). All quantities 
  reported are in lattice units.}
\normalsize\linespread{1.1}
\label{shs42.X}
\end{figure}

\begin{figure}[!tb]
\includegraphics[angle=0,width=8cm]{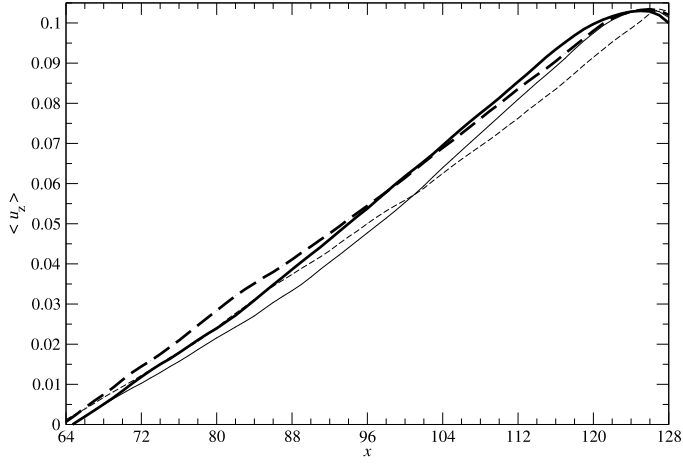}
\caption{\footnotesize\linespread{0.9}
 Spatially averaged velocity component $u_\r{z}$ for the molten gyroid and
 the gyroid mesophases sheared with $U=0.10$, at late times and over the
 $x\ge64$ half of the system. The dashed thin and thick curves
 correspond to the molten gyroid at time steps $\Delta t=9\,000$ and
 13\,000, respectively. The solid thin and thick curves correspond to
 the gyroid at time steps $\Delta t=9\,000$ and 13\,000, respectively. The
 average is over the same two-dimensional domain as described in
 Fig.~\ref{shs42.X}, for each $x$, and its standard error is shown as
 negligible error bars. Note that the velocity shows a maximum located
 from 2 to 4 sites away from the boundary, unlike a simple fluid which
 would display it exactly at the boundary. The value of this maximum
 coincides with the actual velocity at which the BGK relaxation process
 of our LB model is forcing the fluid to move, which needs not coincide with
the
 input parameter $U=0.10$. Note that the inversion in the sign of the stress
 that we reported in Fig.~\ref{shs42.X} occurs precisely for
 $|x-x_0|\le2$, $x_0=0,\,128$ and at (late) times close to and after
 $\Delta t=9\,000$. The behaviour at the other boundary region is
 similar and symmetric to that displayed here. All quantities reported
 are in lattice units.
\normalsize\linespread{1.1}
\label{shs42_53.vel.X}}
\end{figure}

Figure~\ref{shs42.X} shows the profile of the stress, for the sheared gyroid,
along the applied velocity gradient direction. Several curves therein depict
the transport of momentum towards the core (i.e., the plane $x=64$) as the 
strain grows as a function of time. Distinctively, the profiles have
spatial fluctuations, a consequence of the gyroid's convoluted structure whose
interfacial tension locally modifies the viscosity expected for a simple fluid. 
The $u_\r{z}$ component of the velocity field, shown in Fig.~\ref{shs42_53.vel.X}
and averaged in the same way as stated for $\langle -P_\r{xz}\rangle$ in the 
caption of Fig.~\ref{shs42.X}, is however not inhomogeneous but follows a 
transient similar to that expected for a simple fluid: we observe the setting 
up of a steady, smooth and wedge-shaped profile, except at the borders.
Figure~\ref{shs42_53.vel.X} also includes the behaviour of the
averaged velocity profile for the molten gyroid MG at late times, and is seen to
match that of the gyroid G.

\begin{figure}[!htb]
\includegraphics[angle=-90,width=10.5cm]{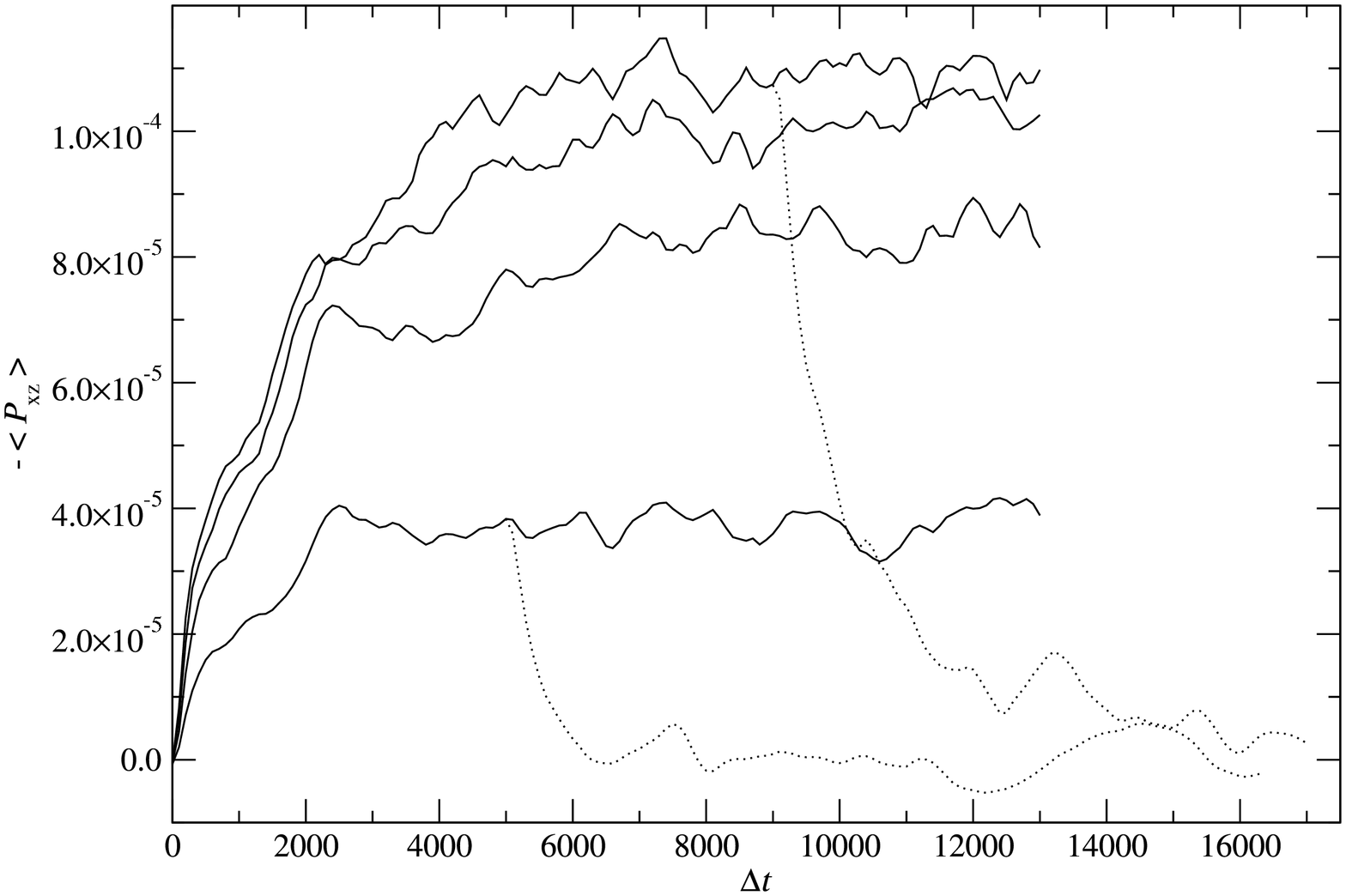}
\caption{\footnotesize\linespread{0.9}
  Temporal evolution of the average shear stress of the gyroid for
  different values of steady shear. The initial condition is the
  same as that mentioned in Fig.~\ref{shs42.X}. The curves, as seen, e.g., at 
  $\Delta t=4000$ from bottom to top, correspond to Lees-Edwards walls 
  moving with speeds \mbox{$U=0.05,\,0.10,\,0.15,\,0.20$}
  (shear rates $S/10^{-3}=0.39,\,0.78,\,1.17,\,1.56$),
  respectively. The dotted curves are the responses after a sudden 
  termination of shear; they are also referred to as the system's 
  relaxation functions for the relevant shear speeds. The average here is 
  in the three-dimensional domain 
  \mbox{$[8,N_\r{x}-8]\times[1,N_\r{y}]\times[16,N_\r{z}-16]$}, where 
  \mbox{$N_\r{x}=N_\r{y}=N_\r{z}=128$} and error bars are the standard 
  error of the average. All quantities reported are in lattice units.
\normalsize\linespread{1.1}
  \label{shs42.Stats}}
\end{figure}

\begin{figure}[!htb]
\includegraphics[angle=-90,width=10.5cm]{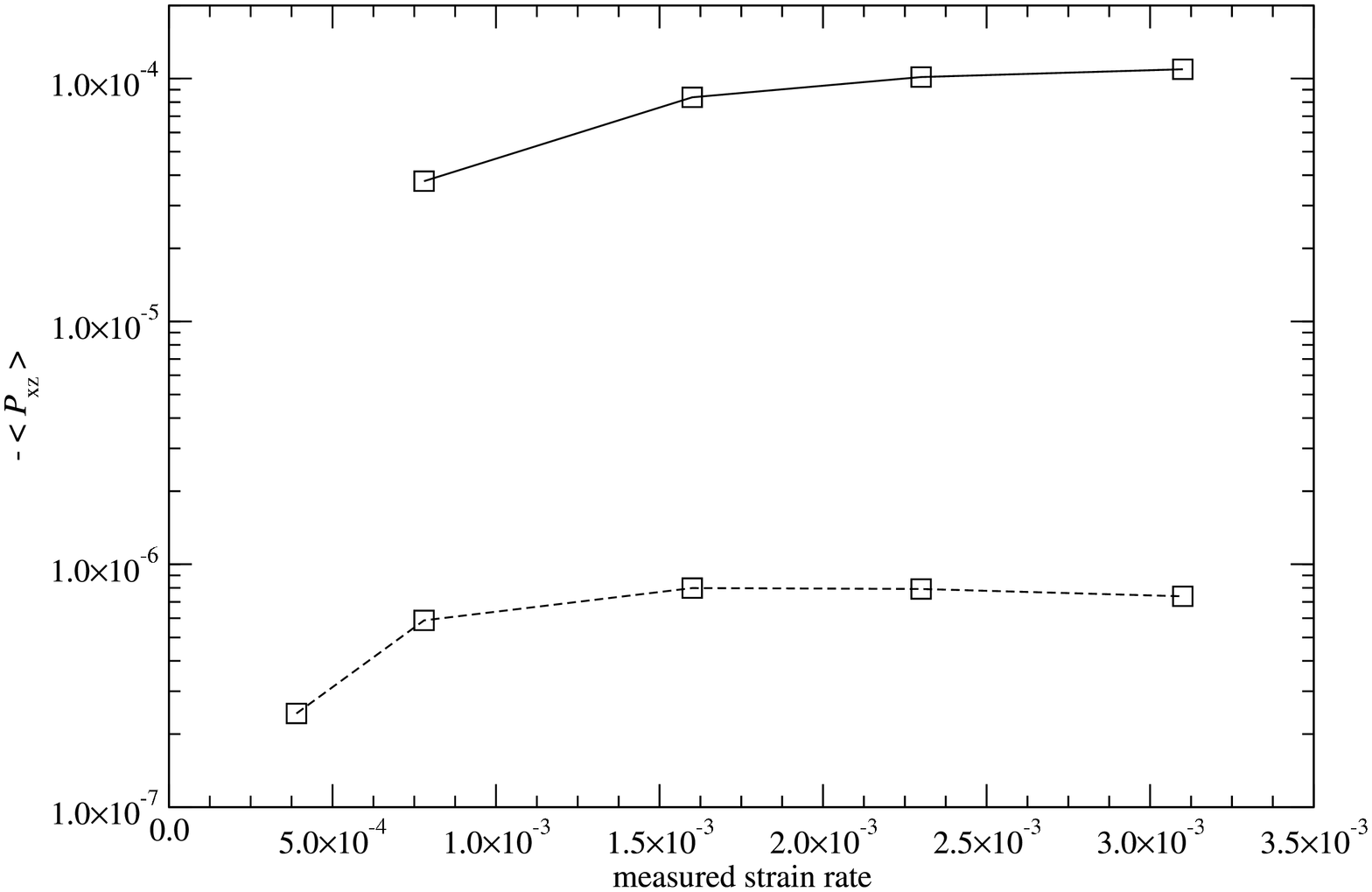}
\caption{\footnotesize\linespread{0.9}
 Both the gyroid (solid line) and the molten gyroid (dashed)
 mesophases exhibit shear thinning. Shown is the stress averaged over 
 the interval $24\,000\le t\le 28\,000$. 
 From the figure it is clear that the gyroid manifests greater
 stiffness than the molten gyroid and its (effective) viscosity drops for
 higher strain rates. All quantities reported are in lattice units.
\normalsize\linespread{1.1}
  \label{nonNewtonian_G_L_3}}
\end{figure}

Remaining with the G mesophase, we show in Fig.~\ref{shs42.Stats} the 
temporal evolution of the stress displayed in Fig.~\ref{shs42.X}; the 
values plotted are averages of the latter on the $8\le x\le N_\r{x}-8=120$ 
interval, which amounts to averaging over the whole lattice except
thin slabs adjacent to the boundaries. In addition to
Fig.~\ref{shs42.X}, we include higher and lower shear velocities,
namely \mbox{$U=0.05,\,0.15,\,0.20$}. Note that the time evolution of
the averaged stress is a succession of peaks and troughs, denoting
successive intervals of yield and recoil, which is a canonical feature of 
viscoelastic behaviour. Were the strain rate at which the gyroid deforms 
coincident with the applied shear rate, these curves would imply shear 
thinning. In fact, while the
increments in applied shear rate between these curves are kept
constant, the increments in the (absolute) values of the stress at
late times do not remain so but decrease. In
Fig.~\ref{nonNewtonian_G_L_3} we show the stress averaged over 
time steps 24\,000  
to 28\,000, plotted against the true strain rate, where the latter was
measured from the linear velocity profile generated at $\Delta
t\ge9\,000$ ($t\ge24\,000$), as displayed in
Fig.~\ref{shs42_53.vel.X}. Figure~\ref{nonNewtonian_G_L_3} clearly
shows shear thinning: the slope, i.e., the effective viscosity 
$\eta^\r{eff}\equiv\partial P_\r{xz}/\partial\dot\gamma$,
decreases with the strain rate. Figure~\ref{nonNewtonian_G_L_3} also 
contains the analogous curve for the molten gyroid, which shows shear 
thinning for the latter at lower 
strain rates than those at which the gyroid does, and at higher
intensity, i.e.
\be
	\frac{\partial\eta^\r{eff}}{\partial\dot\gamma}\Big|_\r{molten}
	<
	\frac{\partial\eta^\r{eff}}{\partial\dot\gamma}\Big|_\r{gyroid}	
	< 0
	\label{CMP_THINNING}\,.
\ee
\noi
This is the first indication of shear thinning reported by means of a 
bottom-up kinetic-theoretic model for fluid flow.

\subsection{Morphological transitions}
\label{MORPH}

\begin{figure*}
\centering
\footnotesize\linespread{0.9}
\begin{tabular}{c}
\mbox{\subfigure[$\Delta t=0$]{
      \includegraphics[angle=0,width=0.315\textwidth]{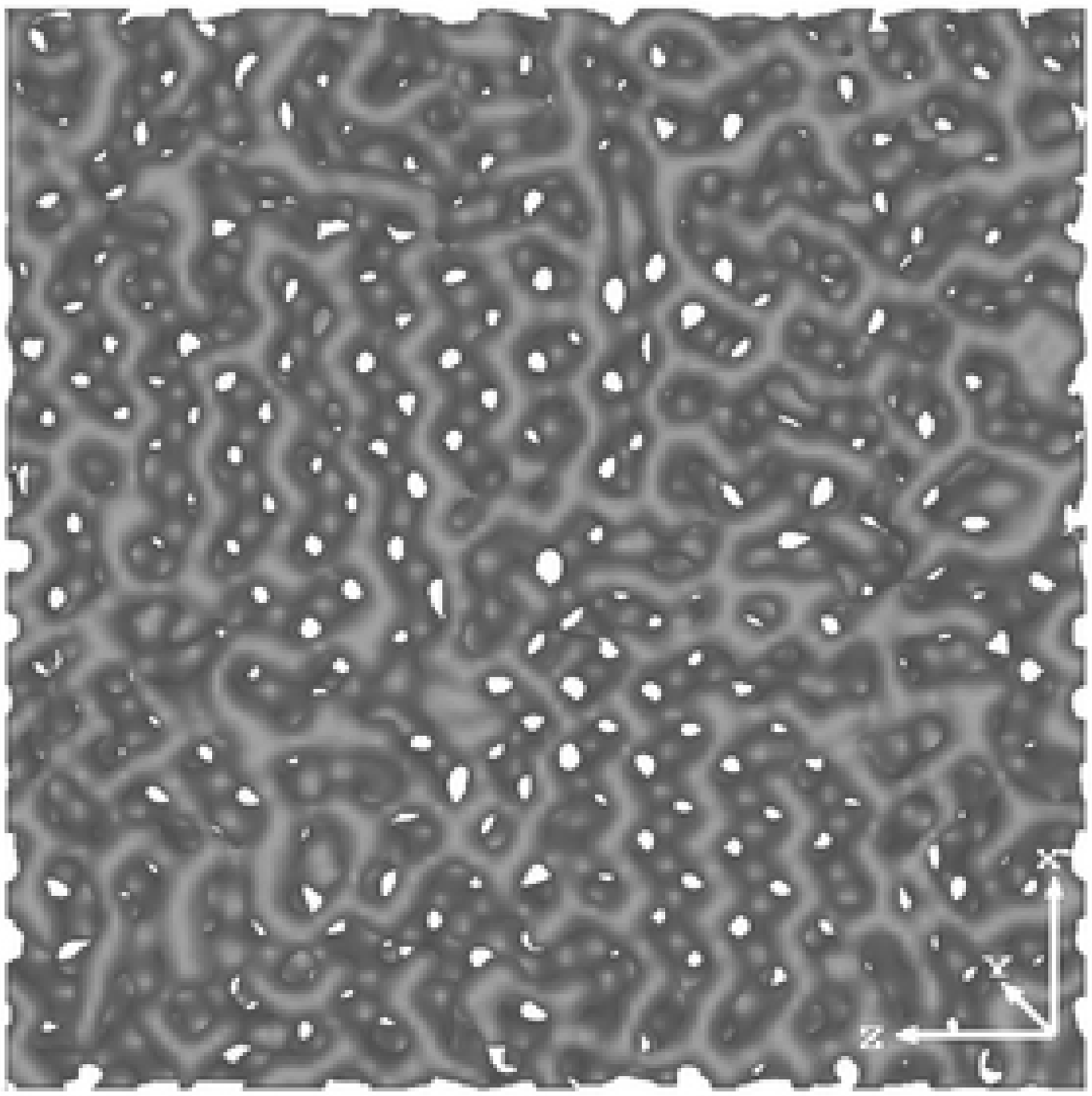}}\quad
      \subfigure[$\Delta t=1\,000$]{
      \includegraphics[angle=0,width=0.32\textwidth]{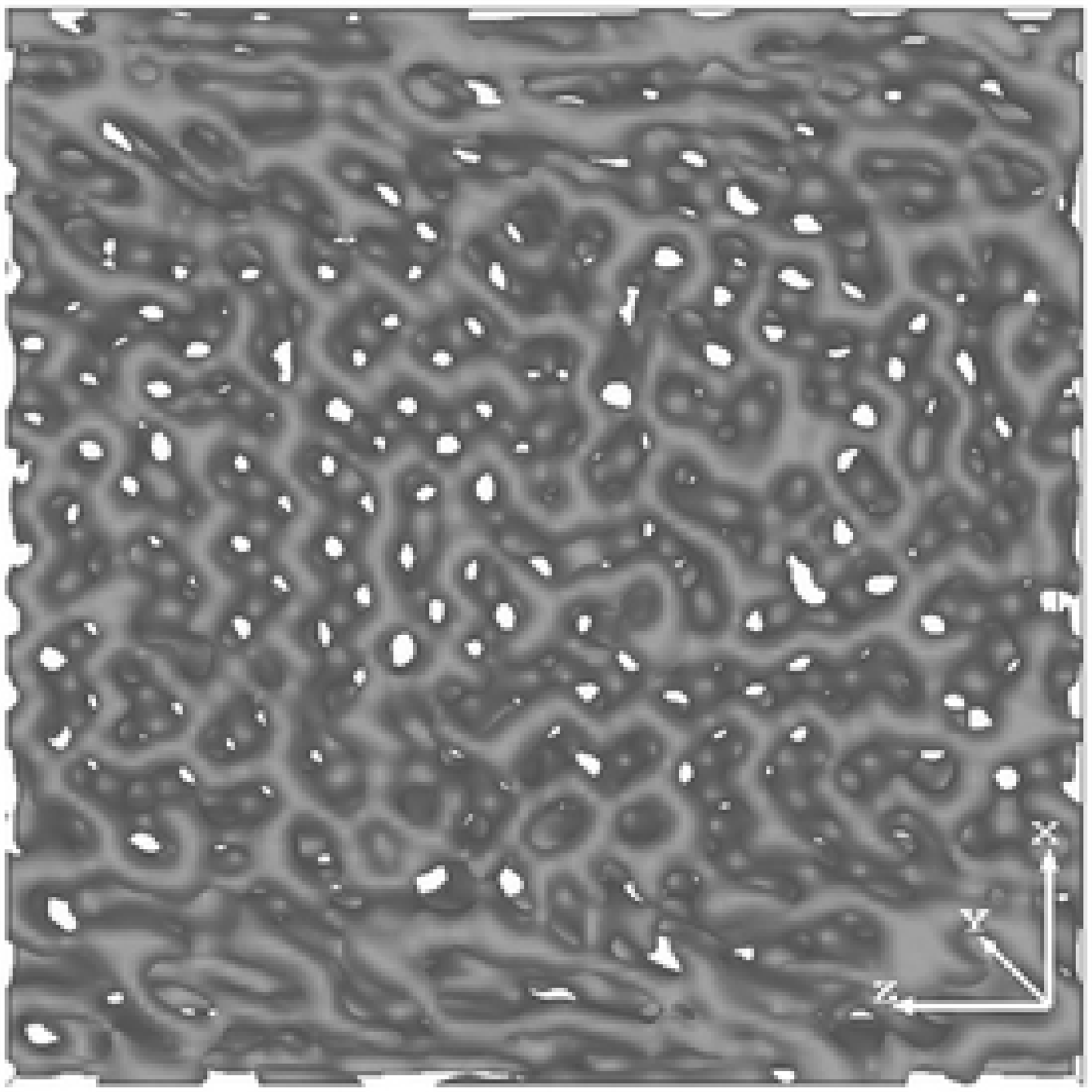}}\quad
      \subfigure[$\Delta t=21\,000$]{
      \includegraphics[angle=0,width=0.312\textwidth]{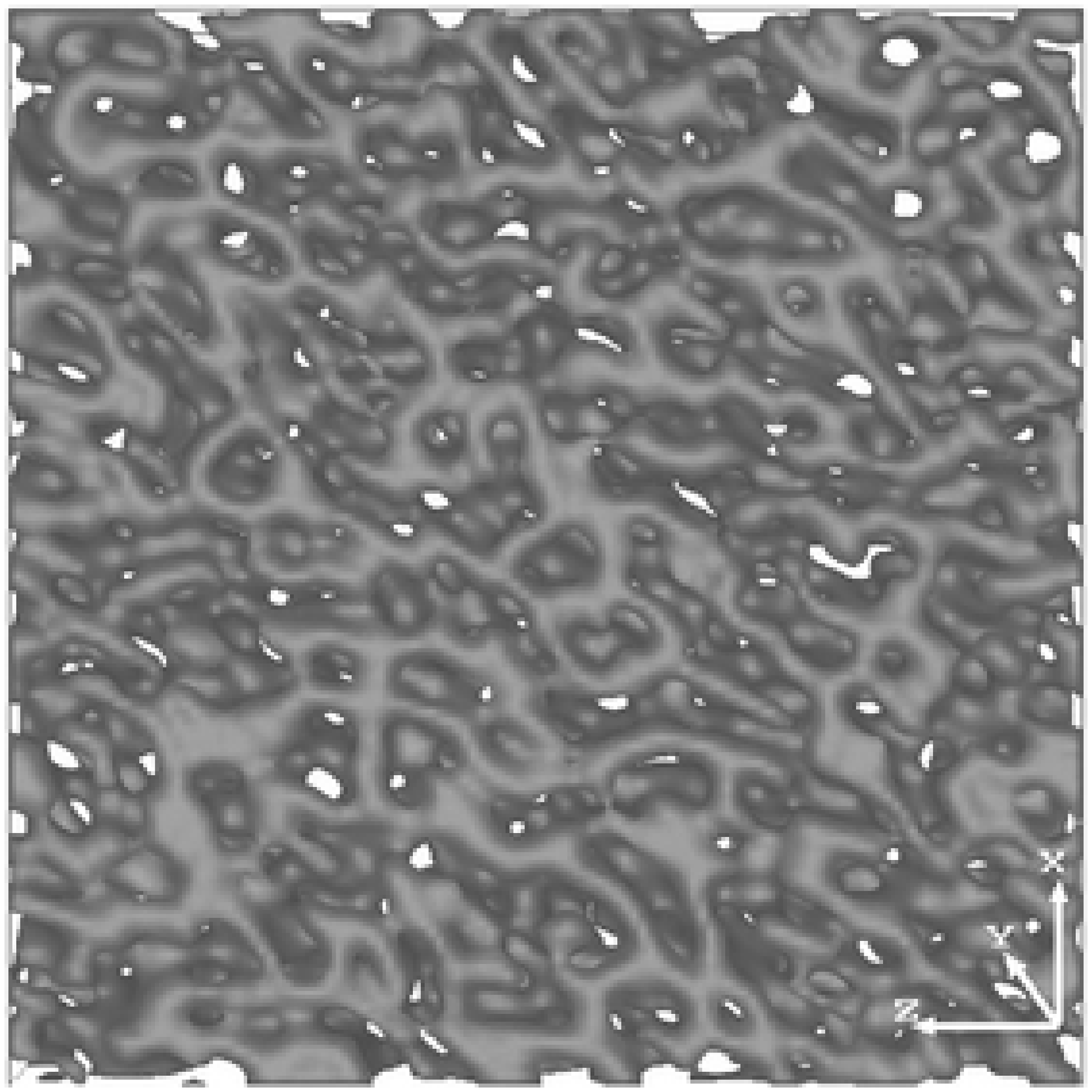}}}
\end{tabular}
\caption{High-density locus of one species (say, oil) in the gyroid
mesophase, before shearing, (a), and at an early, (b), and late time slice, 
(c), after the onset of shear. The shear speed is $U=0.20$. The complementary
immiscible fluid (water) fills the voids with a similar, inter-weaving structure. 
The system is on a \mbox{$128\times128\times128$} lattice, and all figures show 
the subvolume $40\le y\le 52$ and the reference system in use (the $y$-axis is 
perpendicular to the plane of the page). The initial configuration, (a), is a gyroid 
at 15\,000 time steps of self-assembly under periodic boundary conditions. 
These images are volume renderings of the density 
  order parameter, $\phi\equiv\rho^\r{oil}-\rho^\r{water}$; the
  regions visible to the reader are those for which $\phi\ge0.36$
  whilst over the entire fluid $-0.79\le\phi\le0.79$; the regions for which 
$\phi\le -0.36$ (water, not shown) display a similar structure which is complementary
  (interweaving) to the one shown here. All quantities
  reported are in lattice units.}   
\normalsize\linespread{1.1}    
\label{MORPH_G}
\end{figure*}

\begin{figure}
\centering
\footnotesize\linespread{0.9}
\begin{tabular}{c}
  \includegraphics[angle=0,width=7.5cm]{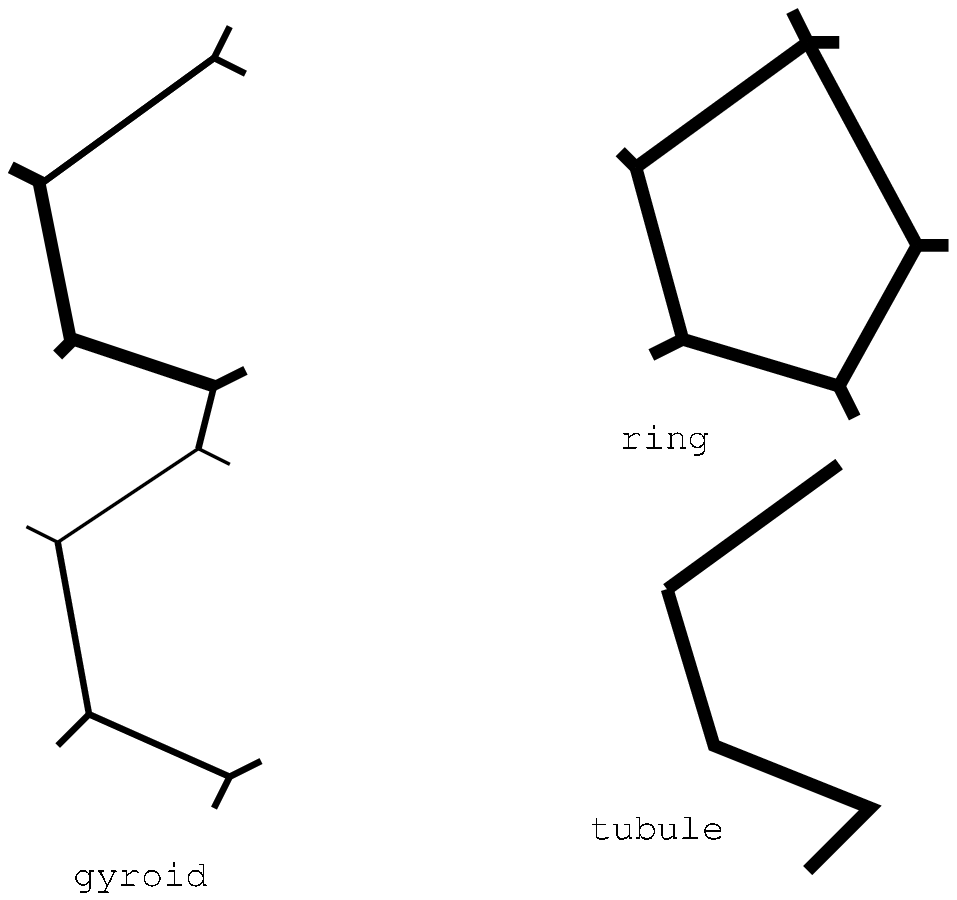}
\end{tabular}
\caption{Schematic representation of the skeleton (locus of highest density) 
of the gyroid mesophase we employ, and two of its structural features before 
and at late times after the onset of steady shear. The thickness provides a
sense of perspective, and represents how close each segment is to the reader;
note that the figures on the right are planar. 
The skeleton denoted by 
`gyroid' depicts a portion of one of the two chiral lattices making up the 
long-range order regions of the gyroid before shear, 
cf.~Fig.~\ref{MORPH_G}a---the 
coordination number is three at each node. In the regions of the gyroid 
containing defects, as well as in most of the sheared mesophase at late times, 
the coordination number can be reduced to two, describing a `tubule'. We also 
show the skeleton of the `ring' structure ubiquitous in the sheared gyroid 
at late times, also present in smaller proportion as a defect in the 
mesophase before the onset of shear. At lower values of density, this ring
appears to be toroidal.} 
\normalsize\linespread{1.1}    
\label{TOROIDAL}
\end{figure}

Figure~\ref{MORPH_G} shows the configuration of the gyroid in the $40\le y\le 52$
slab of the \mbox{$128:128:128$} lattice, before and at late times after applying 
a shear of $U=0.20$. The volume-rendering graphical representation employed~\cite{VTK} 
makes regions where $\phi\ge0.37$ opaque to the lighting rays, assumed to shine
normal to the plane of the text and inwards; since $-0.79\le\phi\le0.79$ over the 
entire system, these regions are the high-density locus 
of one of the species (say, oil). Before shear, the structure contains highly ordered
subvolumes of gyroid symmetry and diagonal length from about 32 to 64 lattice sites,
cf. Fig.~\ref{MORPH_G}(a). This gyroid is hence a collection of subvolumes with a
regular tubular structure making up two three-fold coordinated, interweaving chiral 
lattices of which we depict only one. Since the size of the G unit cell is 
approximately 5 to 6 lattice units, the depth ($y$-dimension) of the slabs shown in 
Fig.~\ref{MORPH_G} is of about two gyroid unit cells. As can be seen in 
Fig.~\ref{MORPH_G}(a), the interfaces between these gyroid subvolumes are defective 
regions where long-range order and symmetry appears to be drastically 
reduced~\cite{P3publon,P3}. Two features characterising them are the spatial variation 
in coordination number and chirality, seen by the presence of elongated tubules 
and toroidal rings, cf.~figure~\ref{TOROIDAL}.

At $\Delta t=21\,000$, which is a late time after the onset of shear and we take as 
steady state, the structure has lost any resemblance with the initial gyroid, 
except for the persistence of the toroidal rings, see Fig.~\ref{MORPH_G}(c), 
which are defects in G. Also, the structure at $\Delta t=21\,000$ is essentially 
the same as that at time step $\Delta t=5\,000$---it is a nonequilibrium steady 
state for at least the previous 16\,000 time steps, a 
time longer than that required for the initial configuration to self-assemble from 
a homogeneous mixture of oil, water and amphiphile. The structure at $\Delta t=21\,000$ 
consists of an irregular network of mainly the same structural elements characterising 
the defective regions before the onset of shear, namely, (a) elongated tubules, with a 
tendency to align along a direction which is a linear combination of directions 
\mbox{$(1,0,0)$} and \mbox{$(0,0,1)$}, and (b) toroidal, ring-like structures. This 
description is, by visual inspection, similar for every subvolume of the lattice 
visualised.

We also looked into the structure of the sheared molten gyroid at late times. In
contradistinction to the gyroid's state at high strain, showing tubules of shape
similar to that depicted in Fig.~\ref{TOROIDAL} and at an angle with the 
$x=\r{const.}$ planes, the highly strained molten gyroid displays tubes which are 
more stretched and aligned along the ${\uvec z}$ direction. The toroidal rings,
also present for the molten gyroid before shear, represent a much smaller volume 
fraction for the sheared molten gyroid than for the sheared gyroid.

\begin{figure*}
\centering
\footnotesize\linespread{0.9}
\begin{tabular}{c}
\mbox{\subfigure[$\sum_{k_\r{y}}S(\vec{k},t),\quad\Delta t=0$]{
      \includegraphics[angle=0,width=0.4\textwidth]
		      {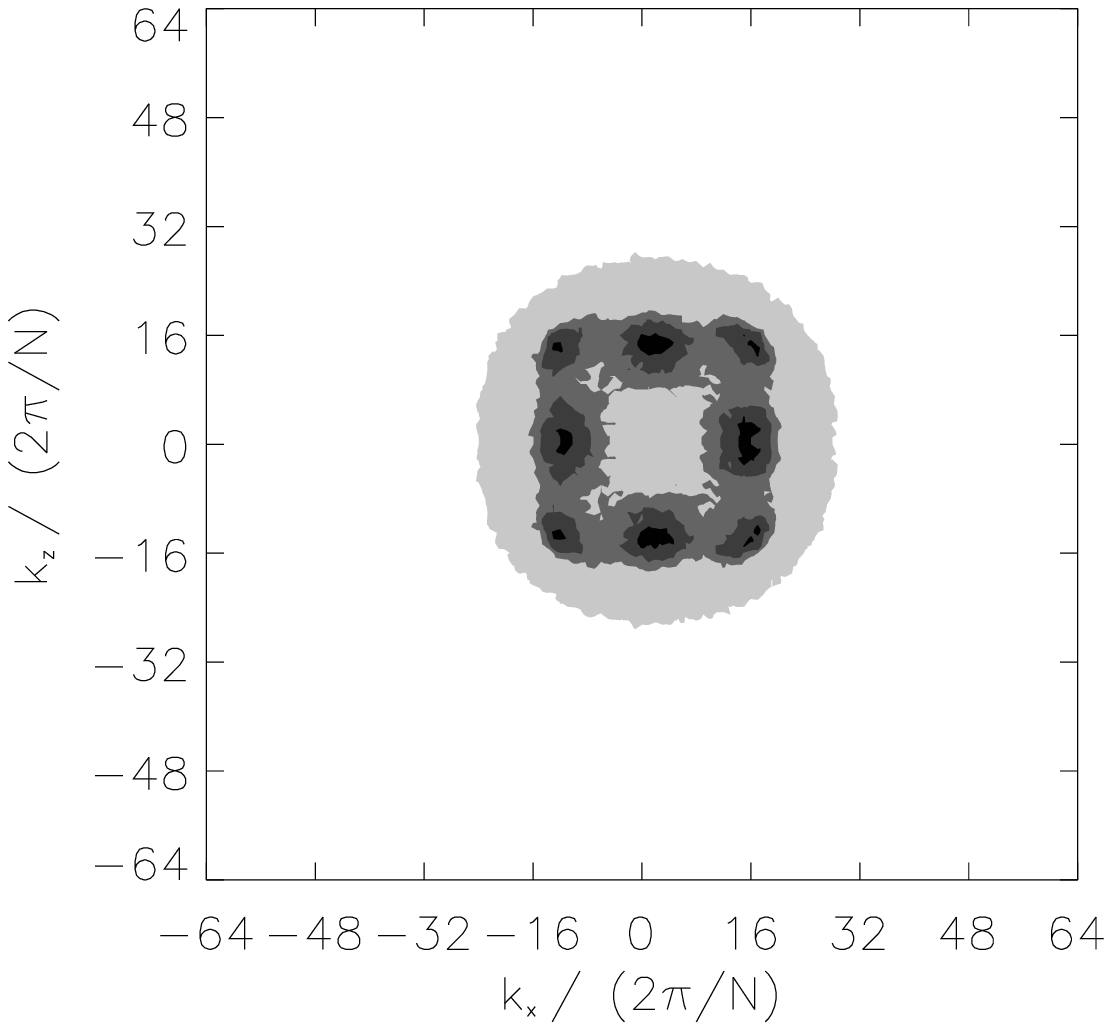}}\quad
      \subfigure[$\sum_{k_\r{y}}S(\vec{k},t),\quad \Delta t=5\,000$]{
      \includegraphics[angle=0,width=0.4\textwidth]
		      {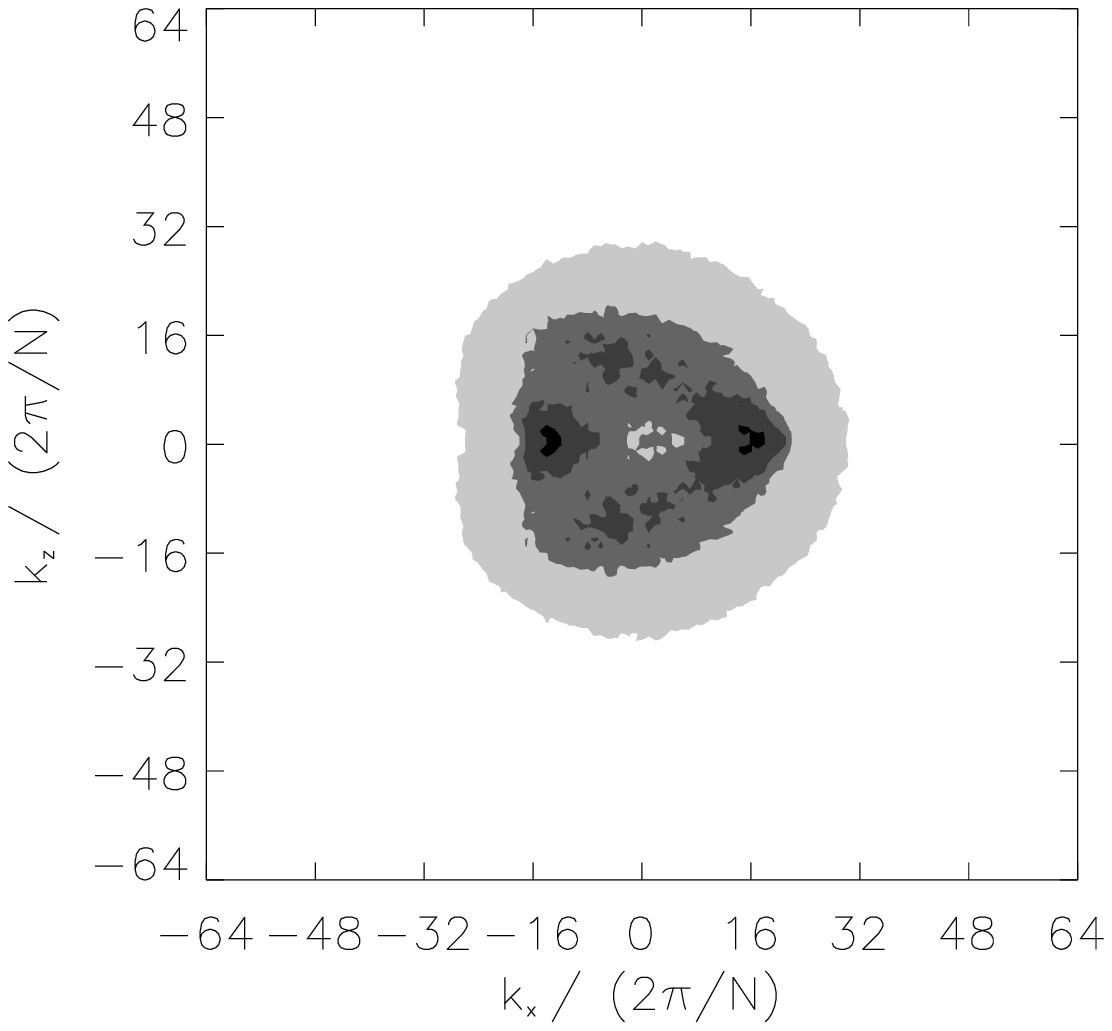}}}\\
\mbox{\subfigure[$\sum_{k_\r{y}}S(\vec{k},t),\quad \Delta t=21\,000$]{
      \includegraphics[angle=0,width=0.4\textwidth]
		      {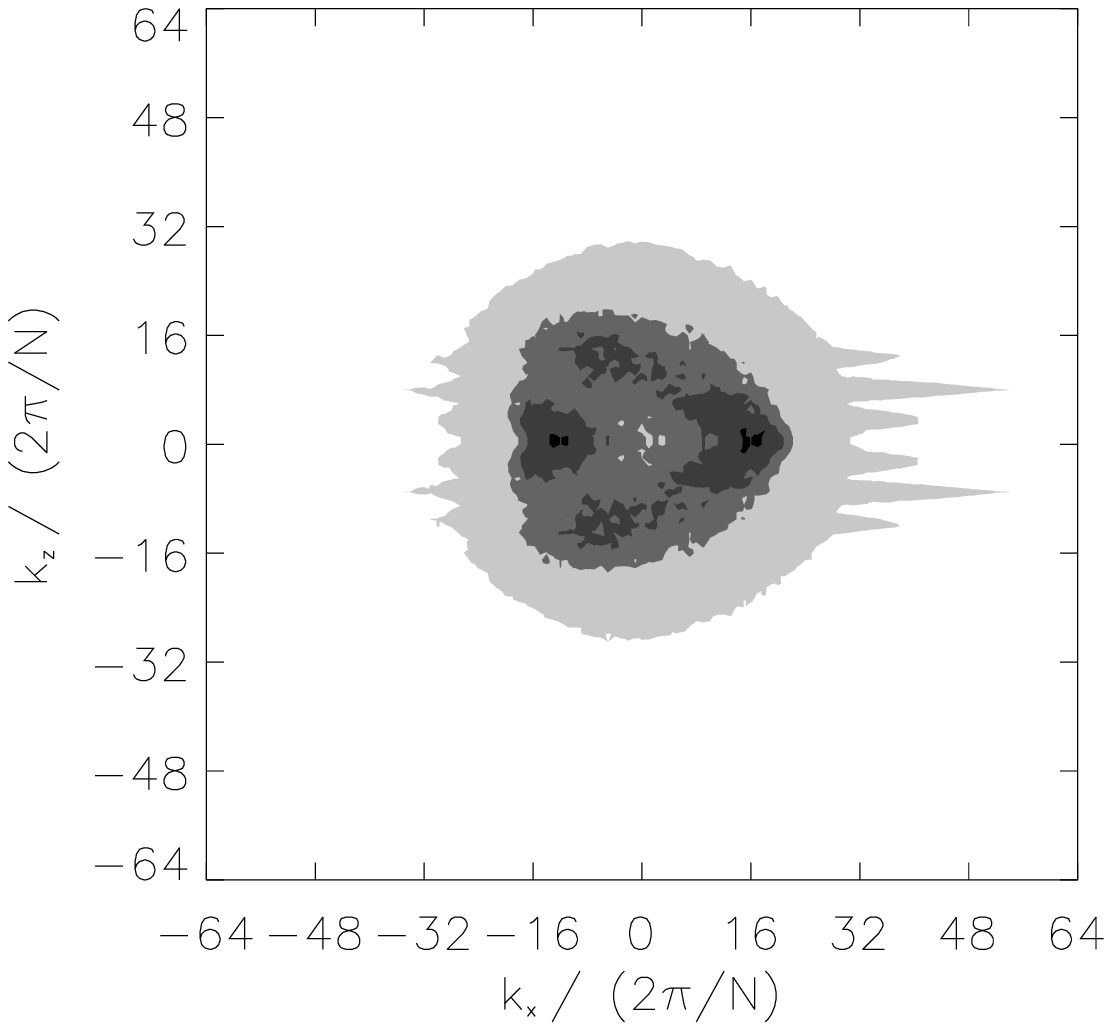}}\quad
      \subfigure[$\sum_{k_\r{z}}S(\vec{k},t),\quad \Delta t=21\,000$]{
      \includegraphics[angle=0,width=0.4\textwidth]
		      {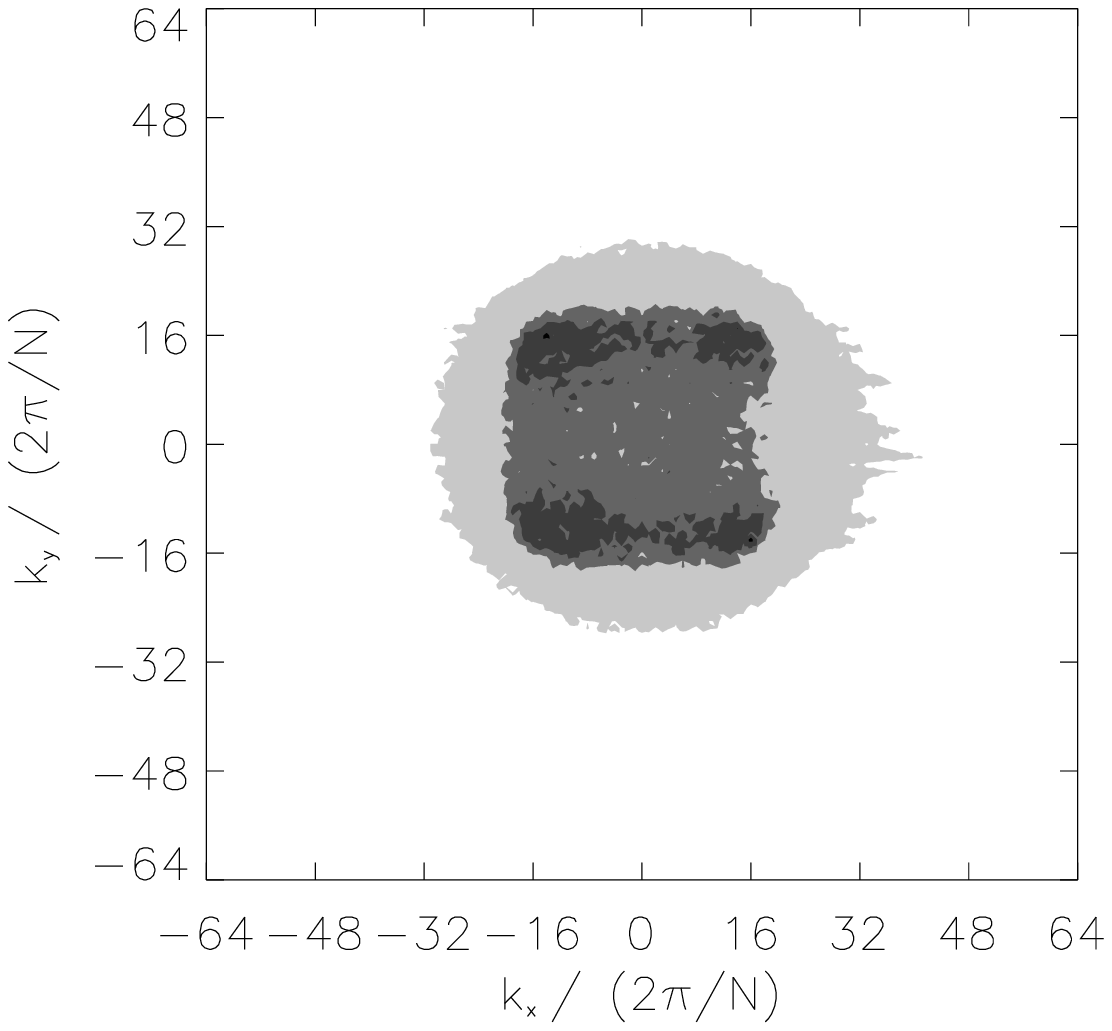}}}
\end{tabular}
\caption{Projected structure function (`scattering pattern') as a 
  function of the time step for the sheared gyroid, as calculated
  using Eq.~(\ref{STRUCT_FUNCT}). Shear velocity is $U=0.20$. 
  (a), (b) and (c) are scattering 
  patterns before shear and at intermediate and late times after the
  onset of shear, respectively, while, for completeness, (d) details 
  the side view of the structure function corresponding to (c). The 
  initial condition for shearing was a gyroid on a $128^3$ lattice at 
  15\,000 time steps of self-assembling. Time steps after the start of 
  shear for these snapshots are indicated below each. Darkness in the 
  greyscale grows with the scattering intensity---filled isocurves correspond 
  to values $S=1,80,200,700$. The spikes are shear-dependent features; see 
  Fig.~\ref{SF_IDEAL_G} and text for discussion.
All quantities reported are in lattice
  units, and $N\equiv N_\r{x}=N_\r{z}$.}  
\normalsize\linespread{1.1}    
\label{SF_G}
\end{figure*}

\begin{figure*}
\centering
\footnotesize\linespread{0.9}
\begin{tabular}{c}
\mbox{\subfigure[$\delta_\r{max}=0$]{
      \includegraphics[angle=0,width=0.32\textwidth]{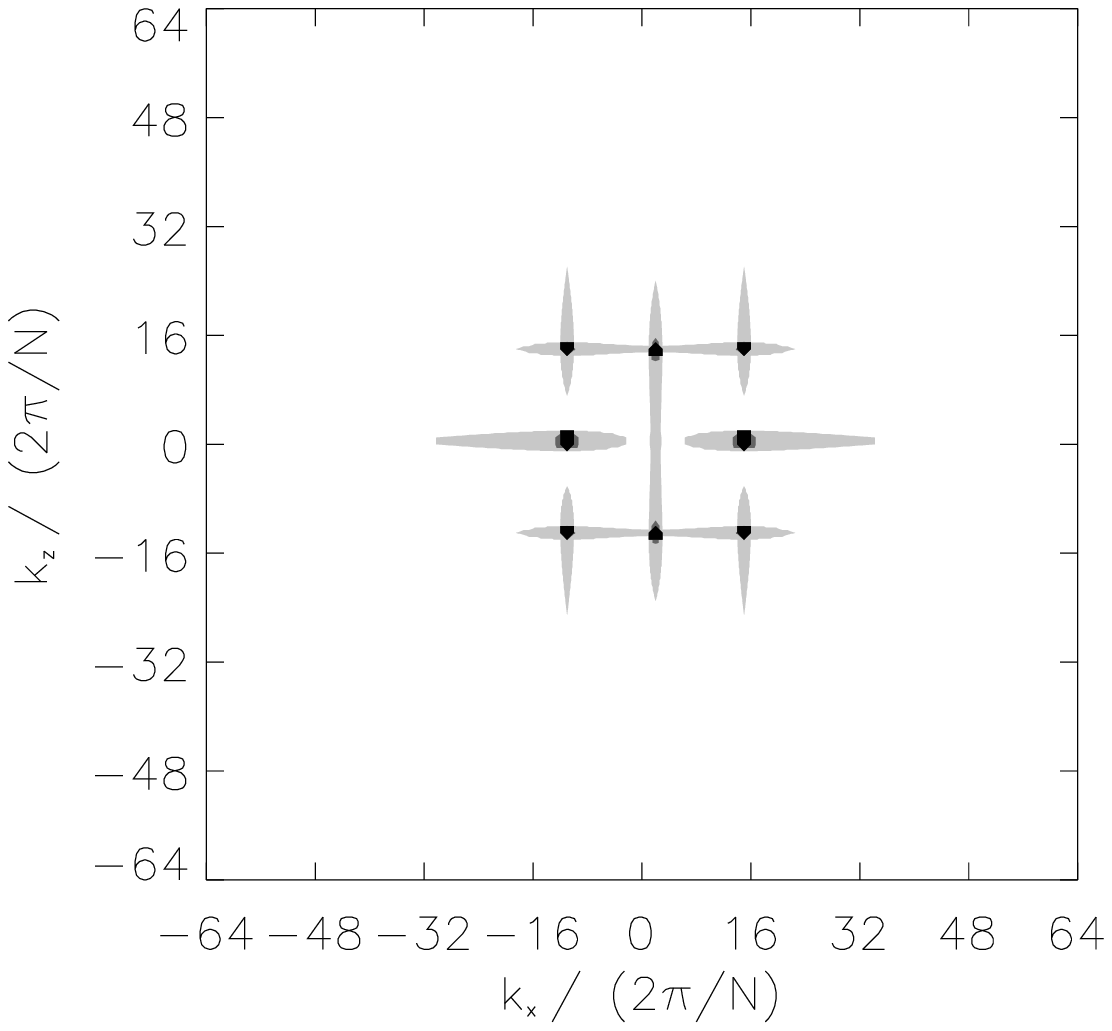}}\quad
      \subfigure[$\delta_\r{max}=8$]{
      \includegraphics[angle=0,width=0.32\textwidth]{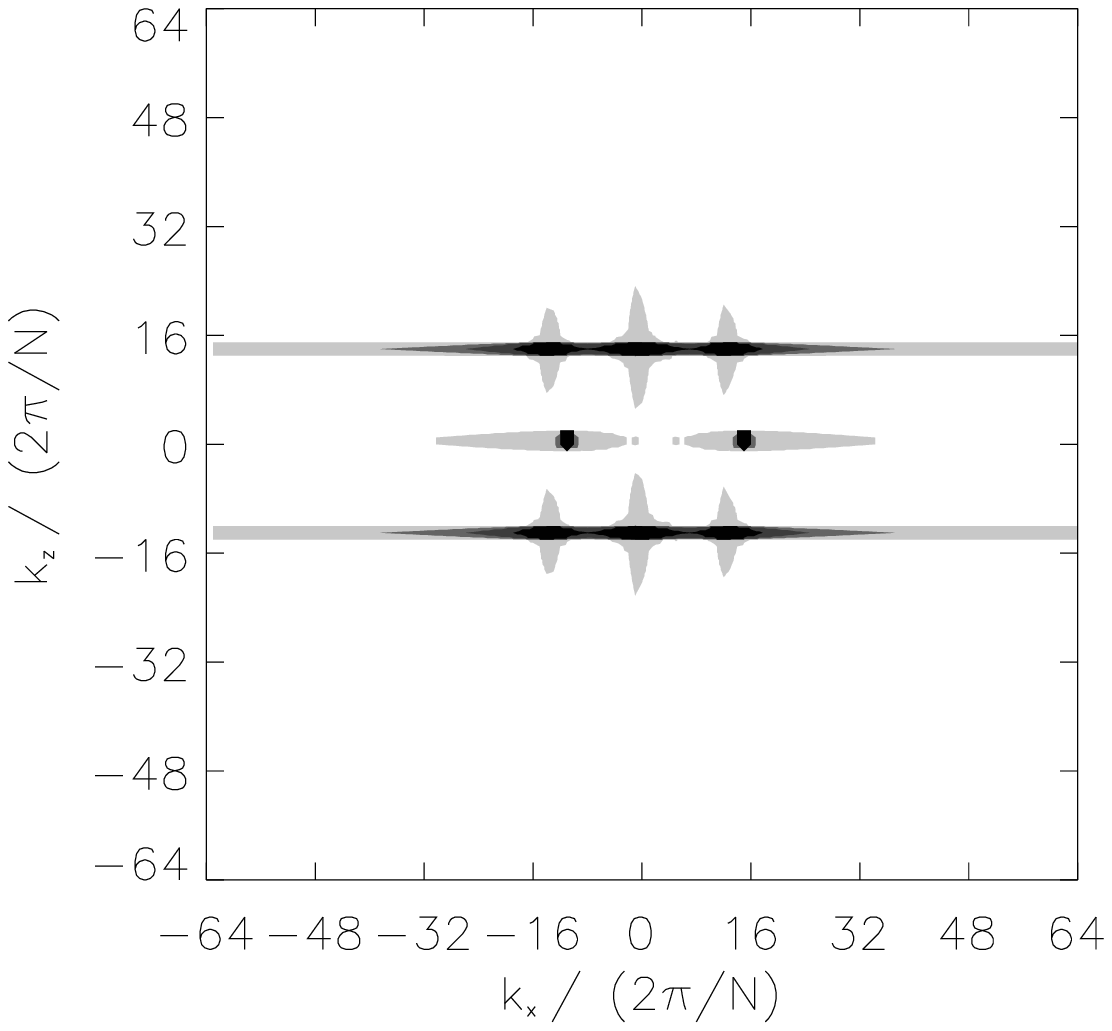}}\quad
      \subfigure[$\delta_\r{max}=16$]{
      \includegraphics[angle=0,width=0.32\textwidth]{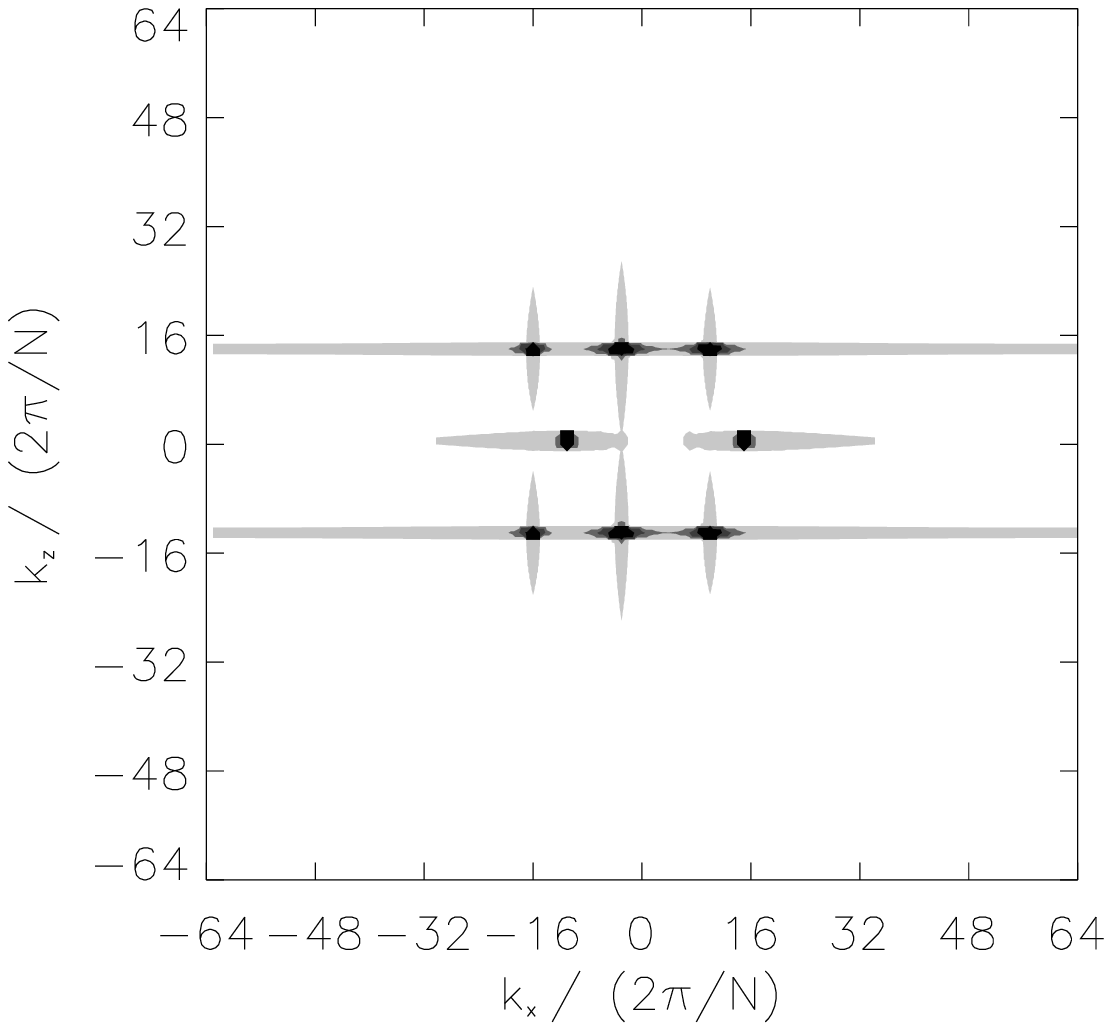}}}
\end{tabular}
\caption{Structure function of the `synthetic gyroid', as calculated using 
  Eq.~(\ref{STRUCT_FUNCT}) on the field $G(\vec{x})$, the expression of the 
  latter being Eq.~(\ref{IDEAL_G}).
  Parameter $\delta_\r{max}$ is the maximum value of the dephase
  $\delta(\vec{x})=(x-N_\r{x}/2)\delta_\r{max}$, which serves to mimic a 
  uniform strain across the structure. The case $\delta_\r{max}=0$ 
  gives an approximation to the Schoen G (or `ideal gyroid') 
  structure. Darkness in the greyscale 
  grows with the scattering intensity, and the filled isocurves shown
correspond
  to $S=1,80,200,700$. For $k_\r{z}\ne0$, the strain shifts
  the pattern leftwards and smear the peaks, while leaving the $k_\r{z}=0$
peaks
  intact. The smearing not being in direct relation to the strain---panel (b) 
  shows more smearing than panel (c)---suggests a similar behaviour for
  the spikes shown in Fig.~\ref{SF_G}(c). The `cardioid' shape reported in 
  Fig.~\ref{SF_G} originates from the facts that the structure undergoes a 
  structural transition (weaking and/or relocation of some of its
$k_\r{z}\ne0$, 
  $k_\r{x}\ne0$ peaks) whilst being sheared with a velocity profile of
positive 
  slope (cf. Fig.~\ref{shs42_53.vel.X}), which orients the `atria' leftwards. 
  All quantities reported are in lattice units, and $N\equiv N_\r{x}=N_\r{z}$.}
 
\normalsize\linespread{1.1}    
\label{SF_IDEAL_G}
\end{figure*}

Figure~\ref{SF_G} shows the summed structure function
$\sum_{k_\r{y}}S({\mathbf k},t)$, or scattering pattern, of the sheared
gyroid mesophase, showing stages of its plastic deformation. Here,
$S({\mathbf k},t)$ is the structure function, computed according to~\cite{P2,P3}
\be
	S({\mathbf k},t)
	\equiv
	\frac{\varsigma}{V}
	\Big|
		\phi_{\mathbf k}^\prime(t)
	\Big|^2					
	\,.\label{STRUCT_FUNCT}
\ee
\noi Here, $\vec{k}$ is the discrete wavevector, $V$ is the lattice volume, 
$\varsigma$ is the unit cell volume for the D3Q25 lattice, and
$\phi_{\mathbf k}^\prime(t)$ is the Fourier transform of the
fluctuations of $\phi$. $S({\mathbf k},t)$ is the Fourier transform of the 
autocorrelation function for the order parameter,
\be
    C_{\phi\phi}(\vec{r},t)
    \equiv
    \langle
        \phi(\xt)\phi(\vec{x+r},t)
    \rangle
\ee
\noi where $\vec{r}$ is a vector lag and the brackets indicate an average over 
the spatial coordinate $\vec{x}$. Figures~\ref{SF_G}(a), (b) and (d) are the 
$xz$ `scattering patterns' of the structures in Fig.~\ref{MORPH_G}, produced 
by summing up the structure function along the $x$ direction. At 
$\Delta t=1\,000$ (not shown), the maximum intensity is
reduced to 29\% of its value at $\Delta t=0$, while there appear horizontal 
`smeared out filaments' of very weak intensity, intrinsically related to the
shearing process, as we shall conclude from Fig.~\ref{SF_IDEAL_G}. At 
$\Delta t=5\,000$ a clear {\em cardioid} shape has developed; the fact that
it persists for the rest of the simulation confirms our observation 
that the system reaches a steady state at time step $\Delta t=5\,000$.
In addition, there is no trace of gyroidal patterns along the $x$-direction.

In order to investigate the origin of the cardioid shape, we computed
the scattering pattern for a `synthetic gyroid',
\bea
   G(\vec{x})
   &\equiv&
   \sin{qx}\cos{qy} + \sin{qy}\cos(qz-\delta(\vec{x})) +    \nonumber\\
   && \sin(qz-\delta(\vec{x}))\cos{qx}
   \,.
   \label{IDEAL_G}
\eea 
\noi where $\delta(\vec{x})=(x-N_\r{x}/2)\delta_\r{max}$ is a 
spatially-varying dephase used
to obtain a linear strain on the morphology (its maximum value,
$\delta_\r{max}$, is reached at the lattice boundaries), and
$q=\r{const.}$ is a wavenumber controlling the size of 
the surface's unit cell. It is known that $G(\vec{x})=0$ for $\delta_\r{max}\equiv0$
is a good approximation to the Schoen ``G'' triply periodic minimal
surface of $Ia\overline{3}d$ cubic symmetry, referred to as `the ideal 
gyroid' hereafter~\cite{G_SYNTH}. Figure~\ref{SF_IDEAL_G} shows the
scattering patterns for the unstrained morphology and for dephases
$\delta_\r{max}=8,\,16$. 

Comparing structure function maps in Figs.~\ref{SF_G} and \ref{SF_IDEAL_G}, 
at the same value of the strain rate, proves useful. For the synthetic gyroid, 
the strain is controlled by the number of unit cells that the dephase causes the 
structure to shift at the lattice boundary, following a linear profile as we 
approach the other boundary going through zero strain at the lattice core. 
For our simulated amphiphilic gyroid, 
however, the strain does not follow a linear profile at early times; instead,
the strain at time $t$ would need to be computed from the integral 
$\frac{1}{N_\r{x}}\int_0^t\int_0^{N_\r{x}}
\r{d}t^\prime\r{d}x \, \partial_x u_\r{z}(\vec{x,t^\prime})$, where 
$t^\prime$ is the time parameter. For the purposes of this paper, however, such
an analysis would be superfluous; in fact, Fig.~\ref{SF_IDEAL_G} already provides us 
with enough information to understand the origin of the cardioid shape. For all 
panels, (a), (b) and (c) therein, the position of the peaks at $k_\r{z}=0$ 
($k_\r{x}/(2\pi/N)\approx-14,\,15$, where $N=128$) are invariant under the 
strain (dephase); not so with the peaks at $k_\r{z}\neq0$, which shift leftwards.
(The shift would be rightwards were $\partial_x u_\r{z}<0$ or $\delta_\r{max}$.) 
The shape of the maps in Figs.~\ref{SF_G}(c) and \ref{SF_G}(d) is that of a 
transformed scattering pattern shifted leftwards. This transformation occurs early, 
between $\Delta t=0$ and $\Delta t=3\,000$, and is characterised by two (strong, 
$S\ge700$) peaks similar to those of the gyroid at $k_\r{x}=0$, and two (weaker, 
$200\le S<700$) peaks at $k_\r{z}=0$.

\section{A simpler case: shearing the lamellar mesophase} 
\label{LAMELLAR}

In the last section we reported on the gyroid displaying higher
shear stress than the molten gyroid. Since the structural transition 
between these two mesophases can be driven by both the amphiphile density
and the inter-amphiphile coupling parameter, as we have reported in the
past~\cite{P3}, our aim in this section is to elucidate 
the role of the amphiphile density alone on the stress response to 
shear; we choose the lamellar mesophase as the subject of study, 
since this is the mesophase with the simplest possible internal interface. 

The initial configuration employed was a cubic $128^3$ lattice with 16
lamellae, stacked perpendicularly to unit vector $\hat{\vec{z}}$. The
lamellae were of alternating, oil-water compositions, separated by a 
thin monolayer of amphiphile;  the thickness
of the immiscible and amphiphilic lamellae were 7 and 1 lattice  
sites, respectively. We populated each lattice site with a value of
density kept constant over the region corresponding to a given
species; each microscopic velocity is assigned the same fraction of
this value. We gave amphiphilic regions the densities
$n^{(0)\r{s}}=0,\,0.80,\,0.95$, and oil and water regions the
densities $n^{(0)\r{r}}=n^{(0)\r{b}}=0.7$. Shear was applied 
perpendicular to the lamellae with the same LEBC employed in
the last section, with speed $U=0.10$.

Before the onset of shear, the case without amphiphile for the lamellar 
initial condition just described is, {\em a priori}, a metastable state in our 
LB model. In fact, the structure has a stationary morphology since 
short-range oil-water forces and the absence of fluctuations maintain 
immiscibility, i.e. a value for the interface steepness, $|\nabla\phi|$; 
however, a large enough perturbation in $\phi$ may allow a fluctuation in
surface tension which drives the entire interface to a radically different 
shape. Another factor disrupting this lamellar morphology is shear, which
may work against the interfacial tension by reducing $|\nabla\phi|$; this
can lead to miscibility ($\phi\equiv0$) for high enough strain rates. Despite 
these arguments, we observed stability for the sheared lamellar mesophase 
without amphiphile, as we report next.

\begin{figure}
\centering
\footnotesize\linespread{0.9}
\begin{tabular}{c}
\includegraphics[angle=0,width=6.5cm]{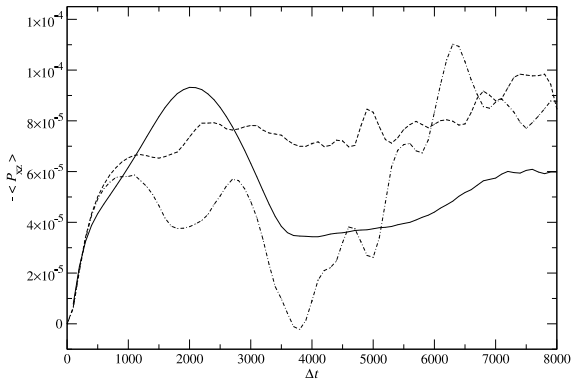}
\end{tabular}
\caption{Temporal evolution of the average shear stress response of a 
  lamellar mesophase at a shear speed of $U=0.10$, for different initial
  amphiphile densities $n^\r{(0)s}$. The solid, dashed, and 
  dash-dotted curves
  correspond, respectively, to \mbox{$n^{(0)\r{s}}=0,\,0.80,\,0.95$}.
  The average is computed over the three-dimensional domain 
  \mbox{$[8,N_\r{x}-8]\times[1,N_\r{y}]\times[16,N_\r{z}-16]$}, where 
  \mbox{$N_\r{x}=N_\r{y}=N_\r{z}=128$} and error bars are not included 
  since they are negligible. All quantities reported are in lattice units.} 
\normalsize\linespread{1.1}    
\label{LAM_STRESS}
\end{figure}

Figure~\ref{LAM_STRESS} shows the stress as measured in the same 
fashion performed on the data plotted in Fig.~\ref{shs42.Stats}, 
for several amphiphile densities. The behaviour observed is diverse. For
zero amphiphile concentration (solid curve), the stress reaches a peak
at early times before it proceeds to a second, lower maximum at late
times, going through a trough at intermediate times due to the fact that 
$|\nabla\phi|$ experiences a transient decrease.

\begin{figure*}
\centering
\footnotesize\linespread{0.9}
\begin{tabular}{c}
\mbox{\subfigure[$n^{(0)\r{s}}=0$]{
      \includegraphics[angle=0,width=0.317\textwidth]{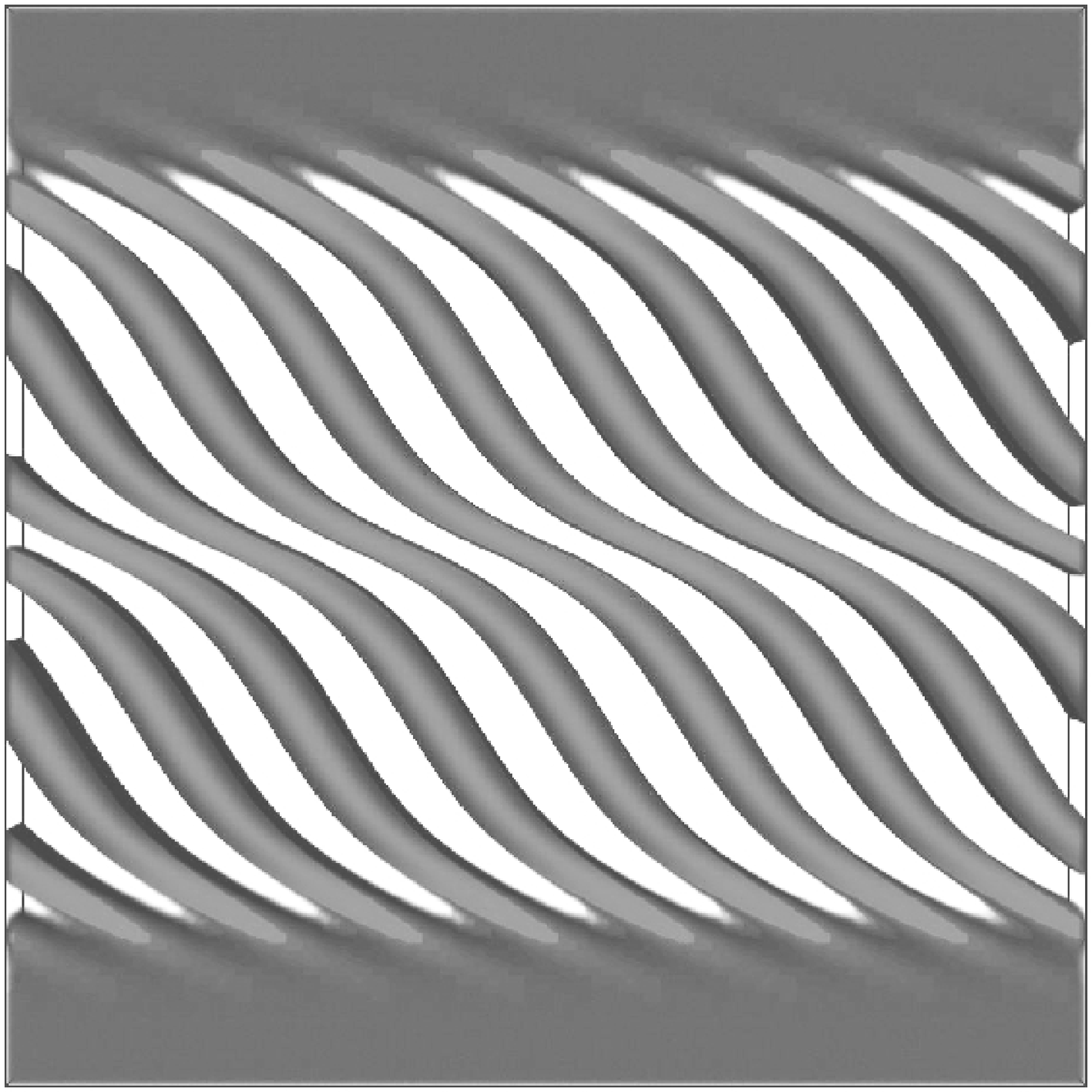}}\quad
      \subfigure[$n^{(0)\r{s}}=0.80$]{
      \includegraphics[angle=0,width=0.313\textwidth]{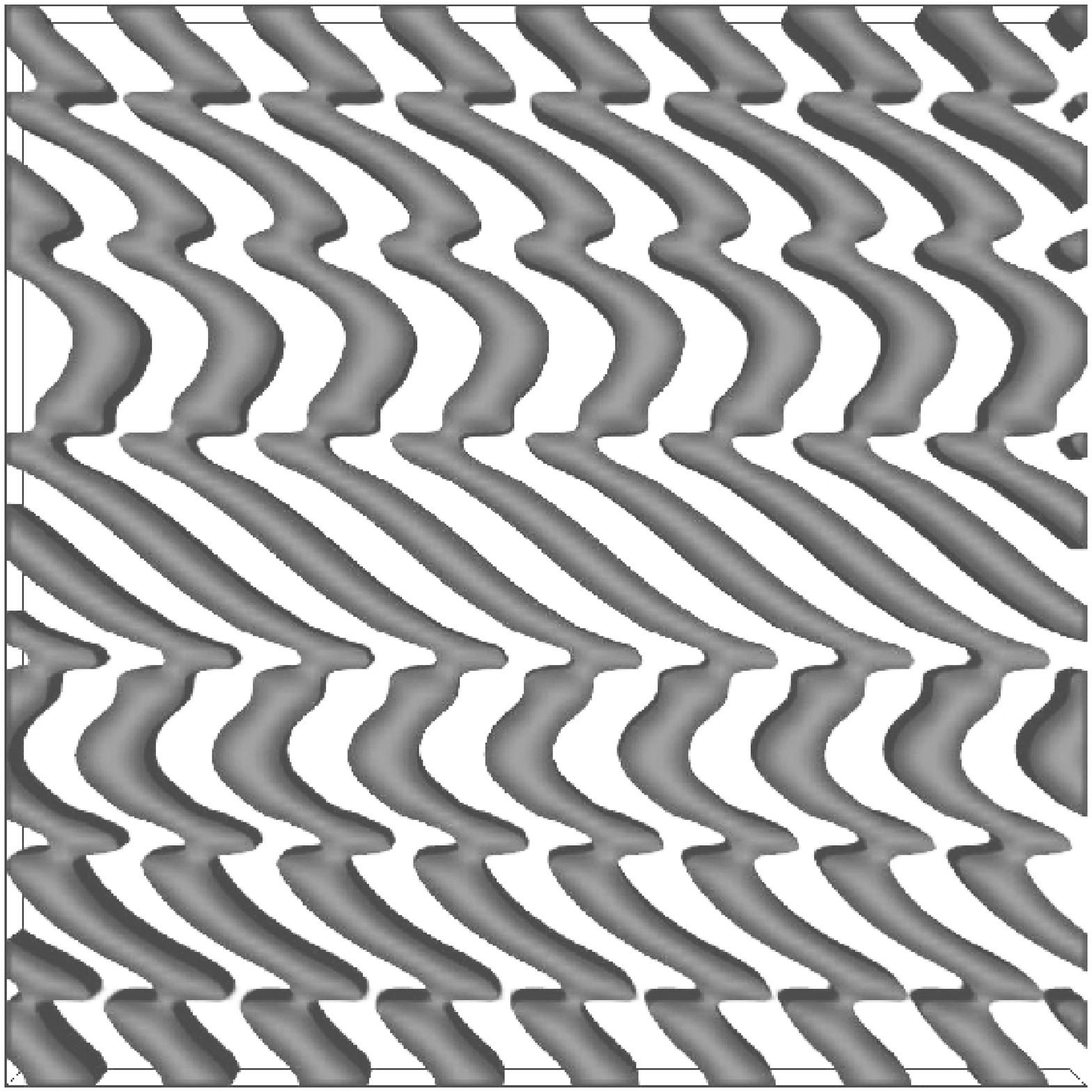}}\quad
      \subfigure[$n^{(0)\r{s}}=0.95$]{
      \includegraphics[angle=0,width=0.32\textwidth]{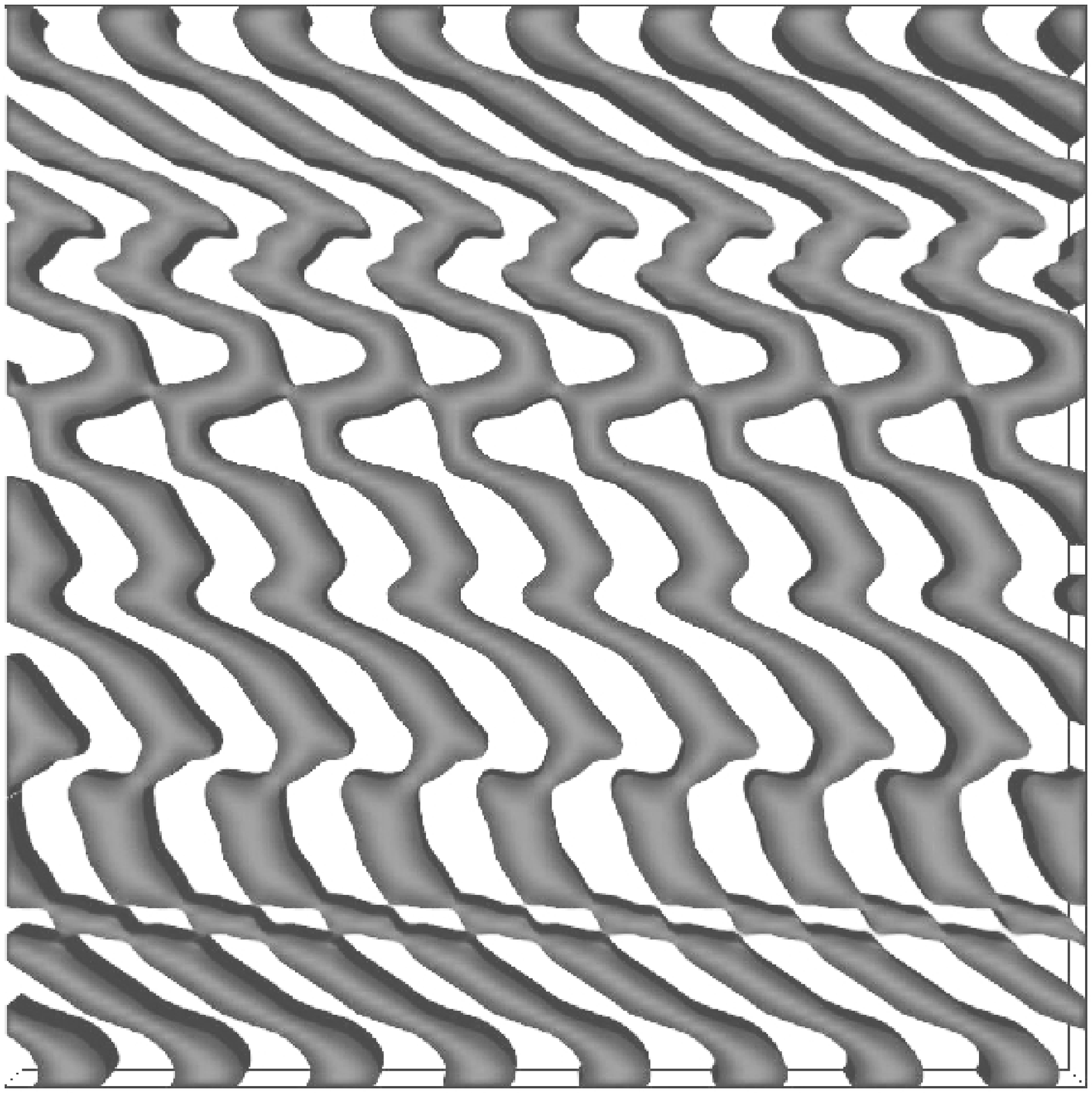}}}
\end{tabular}
\caption{Slabs $0\le y\le 8$ of the order parameter $\phi$ for sheared lamellar
mesophases corresponding to increasing amphiphile density, $n^\r{(0)s}$, as
indicated below each relevant panel, at time step $\Delta t=8\,000$ after 
the onset of shear and for shear velocity $U=0.10$. The coordinate system is 
the same as that in Fig.~\ref{MORPH_G}. In panel (a), the regions opaque to 
incident (volume rendering) light are those for which $\phi\ge0.18$, where
$|\phi|\le0.36$ across the system. In panel (b), the opaque regions are
those for which $\phi\ge0.22$, where $|\phi|\le0.45$ across the system. 
In panel (c), the opaque regions are those for which $\phi\ge0.24$,
where $|\phi|\le0.48$ across the system. It is worth noting that the
surfactantless case, (a), exhibits a curved interface. The amphiphilic
cases, (b) and (c), display the formation of irregularities
in the interface and nodal planes, as a result of the
inter-amphiphile interaction. All configurations have translational
symmetry along the $y$-axis. All quantities reported are in lattice
units.}  
\normalsize\linespread{1.1}    
\label{LAM_COLOUR}
\end{figure*}

The high-density regions of one of the immiscible species (say, oil) 
is shown in Fig.~\ref{LAM_COLOUR}(a) at late times, $\Delta t=8\,000$;
these are representative of the shape of the oil-water interface. Away 
from the boundaries ($x=0,\,128$), there is a large interfacial area
with zero curvature, where we define the curvature as 
$H\equiv\partial^2_{zz} x_\phi(z)$, $x_\phi(z)$ being the curve resulting 
from projecting the $\phi=0.18$ surface onto the $xz$ plane. Curiously, 
we observe three changes of curvature as we follow the curve $x_\phi(z)$ 
for $y=\r{const.}$, namely, $H<0, H>0, H<0, H>0$; instead, we would have 
expected the steady, late-time configuration for the sheared lamellar 
mesophase to rather minimise the interfacial area, leaving only one
inflexion point. For fluid regimes under conditions of local thermodynamic
equilibrium, we can think of $H^2$ as an interfacial free energy 
density associated with the curvature~\cite{LIPOWSKY}; in this case, we would have 
expected the steady, late-time configuration to also minimise the interfacial 
free energy.

The stress curve corresponding to $n^{(0)\r{s}}=0.80$ (cf.~Fig.~\ref{LAM_STRESS})
shows the absence of large troughs, as occurs for 
the $n^{(0)\r{s}}=0$ case, despite the fact that interfacial tension is 
drastically reduced by the presence of the amphiphile. In addition, 
the stress grows at late times to higher values than those achieved by the
$n^{(0)\r{s}}=0$ case. The late-time order-parameter 
configuration is displayed in Fig.~\ref{LAM_COLOUR}(b), showing a rich 
interfacial pattern. Using the same arguments as those of the last paragraph, 
this structure could be characterised by a higher curvature energy, 
$\int\r{d}^2x \,H^2$, where $\r{d}^2x$ is a measure on the oil-water 
interface, and $H$ is now defined as the inverse radius of curvature, 
parametrised on the arclength, $s$. Figure~\ref{LAM_COLOUR}(b) shows 
similar regions of high 
curvature at an equal distance from the shearing walls, where 
$u_\r{z}=\r{const.}$, which we shall call {\em nodal} planes. Also note 
that the interface, as approximately depicted by the boundary of the 
$\phi\ge0.22$ volume, joins the lattice boundary at an angle close to 90 
degrees. 

The stress curve for the $n^{(0)\r{s}}=0.95$ case shows a dramatically
different situation for the first 5\,000 time steps: the presence of a
trough, deeper than that present for the $n^{(0)\r{s}}=0$ density. After
that, there appears a shoot-off whereby the stress rapidly grows and 
equals the late time value achieved in the $n^{(0)\r{s}}=0.80$ case, while
the order-parameter displays a configuration analogous to the 
$n^{(0)\r{s}}=0.80$ case, cf.~Fig.~\ref{LAM_COLOUR}(c). By looking at the 
amphiphile density field, $\rho^\r{s}(\vec{x})$, for the case 
$n^{(0)\r{s}}=0.95$, we observed that the high curvature regions arise 
close to the boundaries first ($\Delta t <1000$), and then rapidly move 
away from them as the strain progresses.

\section{Conclusions}
\label{CONCL}

In this paper we have reported on the shear stress response of two gyroidal 
cubic amphiphilic mesophases previously self-assembled using the same 
bottom-up LB model we employ here, namely, the gyroid {\em per se}, G,
which shows high crystallinity at late self-assembly times, and the molten 
gyroid, MG, endowed with shorter-range order and located within the sponge-gyroid 
lyotropic structural transition~\cite{P3}. Shear was imposed via sliding
periodic (Lees-Edwards) boundary conditions, and we investigated the 
response to several values of the strain rate. In addition, in order to 
investigate the dependence of the shear stress on the amphiphile density, 
we also sheared a lamellar mesophase, of much simpler morphology than the
gyroidal mesophases.

We found that the gyroidal mesophases exhibit shear thinning, more 
pronounced and at lower strain rates for the MG mesophase than the G mesophase. 
In other words, momentum introduced into the system due to shear is 
transported more easily for the mesophase containing more amphiphile, with
longer-range ordering, i.e. the effective viscosity is higher for the G
mesophase.

We also found a shear-induced transition from an initial gyroidal morphology 
(G and MG) to a mesophase characterised by coexisting elongated tubules and 
toroidal, ring-like structures. The features of this mesophase is in contrast 
to those of the mesophase reported by Zvelindovsky et al. using free-energy 
Langevin-diffusion methods by shearing a bicontinuous structure reminiscent of 
a molten gyroid~\cite{ZVELIN3}. The structure they found is of a shorter-range 
ordering than that of the MG mesophase described here, 
and the high-strain structure consists of coexisting lamellae and hexagonally 
packed tubes elongated along the direction of the imposed shear velocity. Our 
sheared mesophases also show enlongated tubes along this direction, but 
the structure is far more complicated than that found by Zvelindovsky et al.
in that it exhibits remnant toroidal rings and `hard shoulders' reminiscent of
gyroidal skeletons; hexagonal packing and coexisting lamellae are, on the other
hand, absent.

The shear performs a plastic deformation which effectively breaks the links
of the gyroidal skeleton; this happens as these links interpose an (oil-water) interface 
whose normal, $\vec{n}$, is parallel or anti-parallel to the flow streamlines, 
$\vec{u}$. In other words, shear effectively applies a `mixing' force which is 
in competition with the inter-particle forces keeping the mesophase in place, 
namely, those controlled by coupling parameters $g_\r{br}$, $g_\r{bs}$ and 
$g_\r{ss}$. Our hypothesis is that adsorbed dipoles sitting on interfacial 
regions at an angle $\theta\equiv\angle{(\vec{u,n})}$ other than $\theta=0,\pi$ 
require more work from the shear forces to be drawn away from the interface than 
those regions on which the streamline impinges normally, since the mixing force
goes as \mbox{$\cos\theta$}. In particular, since the mixing force vanishes for 
$\theta=\pi/2$, considerably longer interfaces can survive the flow---shear 
induces a preferential direction along which the long-range order present 
before the onset of shear is not reduced. This explains not only the formation 
of the elongated tubules but also their reconnection (increase in coordination 
number). In fact, the toroidal, ring-like structures are not only vestigial 
gyroid defects which have survived the gradient $\nabla\vec{u}$, but are also 
born anew as a result from reconnections. It is relevant to point out that 
Padding and Boek, using a coarse-grained molecular dynamics model for wormlike 
micelles, reported on the formation of rings when applying steady shear to a 
wormlike micellar mesophase~\cite{BOEK}---this is an `amphiphile-in-water' binary
mesophase, in contrast to the `oil-amphiphile-water' ternary mesophases we study 
in this paper. 

By applying shear to a lamellar mesophase we found that the presence of 
amphiphile on the oil-water interface of the mesophase causes the interface 
to fold into a wealth of structures with a (discrete) translational 
symmetry on planes equidistant to the shearing walls and along the direction 
of the shear velocity. In other words, the inter-amphiphile force couples the 
adsorbed amphiphilic dipoles so that the interface locally increases its 
curvature energy density. It is worth investigating whether this local
increase is due to the amphiphile being incapable of sustaining
interfacial regions of low curvature under shear, i.e. whether the `breaking' 
mechanism induced by shear is counteracted by regions of high curvature energy 
density. Regarding the shear stress, our
amphiphile-containing lamellae responded with higher stress at late
times than those without amphiphile. This contrasts with the results found for
the gyroidal mesophases, and lets us conclude that it is the
gyroid's cubic morphology that allows this structure to be stiffer.
Understanding the behaviour of the lamellar mesophase under shear requires
the study of amphiphile self-assembly under shear, including in and out of plane
amphiphilic and associated Marangoni currents, and their coupling 
to the imposed flow.

\section{Acknowledgments}
We thank Dr.~Rafael Delgado-Buscalioni, Prof.~Antonio Coniglio and
Prof.~Francesco Sciortino for enlightening discussions. We acknowledge 
Iain Murray, Elena Breitmoser and Jonathan Chin for their involvement in algorithm 
implementation and optimisation within the RealityGrid and TeraGyroid projects. 
This work was supported by the UK EPSRC under grant RealityGrid GR/R67699 which
also provided access to a 512-processor SGI Origin3800 platform at
Computer Services for Academic Research (CSAR), Manchester Computing,
UK. We also thank the Higher Education Funding Council for England
(HEFCE) for our on-site 16-node SGI Onyx2 graphical supercomputer.



\end{document}